\documentstyle[11pt,aas2pp4,psfig]{article}
\def\hi{\ion{H}{1}}
\def\hii{\ion{H}{2}}
\def\hei{\ion{He}{1}}
\def\heii{\ion{He}{2}}

\def\ciii{\ion{C}{3}}
\def\civ{\ion{C}{4}}
\def\niii{\ion{N}{3}}

\def\ga{\ifmmode  \gtrsim \else $\gtrsim$\fi}
\def\la{\ifmmode  \lesssim \else $\lesssim$\fi}
\def\etal{et al.~}
\def\halpha{\ifmmode {\rm H{\alpha}} \else $\rm H{\alpha}$\fi}
\def\hbeta{\ifmmode {\rm H{\beta}} \else $\rm H{\beta}$\fi}
\def\hgamma{\ifmmode {\rm H{\gamma}} \else $\rm H{\gamma}$\fi}
\def\masl{\ifmmode  {\rm M_{\sun}yr^{-1}} \else ${\rm M_{\sun}yr^{-1}}$\fi}

\def\mdot{\ifmmode  \dot{M} \else $\dot{M}$\fi}
\def\msun{\ifmmode M_{\odot} \else $M_{\odot}$\fi}
\def\vinf{\ifmmode v_{\infty} \else $v_{\infty}$\fi}
\def\teff{\ifmmode T_{\rm eff} \else $T_{\rm eff}$\fi}
\def\logg{\ifmmode \log g \else $\log g$\fi}
\def\loggeff{\ifmmode \log g_{\rm eff} \else $\log g_{\rm eff}$\fi}
\def\rstar{\ifmmode R_{\star} \else $R_{\star}$\fi}
\def\lstar{\ifmmode L_{\star} \else $L_{\star}$\fi}
\def\mstar{\ifmmode M_{\star} \else $M_{\star}$\fi}
\def\rsun{\ifmmode R_{\odot} \else $R_{\odot}$\fi}
\def\lsun{\ifmmode L_{\odot} \else $L_{\odot}$\fi}
\def\zsun{\ifmmode Z_{\odot} \else $Z_{\odot}$\fi}
\def\12c16o{$^{12}{\rm C}\left(\alpha,\gamma\right)^{16}{\rm O}$}
\def\kms{\ifmmode {\rm km \;s^{-1}} \else $\rm km \;s^{-1}$\fi}
\def\nlte{non--LTE}

\begin{document}

\title{New Models for Wolf-Rayet and O Star Populations in Young
Starbursts}
\author{Daniel Schaerer}
\affil{Space Telescope Science Institute, 3700 San Martin Drive, Baltimore,
	MD 21218, USA}
\authoremail{schaerer@stsci.edu}

\and

\author{William D. Vacca\altaffilmark{3}}
\affil{Institute for Astronomy, Honolulu, HI 96822, USA}
\altaffiltext{3}{Beatrice Watson Parrent Fellow}
\authoremail{vacca@athena.ifa.hawaii.edu}

\begin{abstract}
Using the latest stellar evolution models, theoretical stellar spectra,
and a compilation of observed emission line strengths from Wolf-Rayet
(WR) stars, we construct evolutionary synthesis models for young
starbursts. We explicitly distinguish between the various WR subtypes
(WN, WC, WO), whose relative frequency is a strong function of metallicity,
and we treat O and Of stars separately.

We calculate the numbers of O and WR stars produced during a
starburst and provide detailed predictions of UV and optical emission
line strengths for both the WR stellar lines and the major nebular
hydrogen and helium emission lines, as a function of several input
parameters related to the starburst episode.  
We also derive the theoretical frequency of WR-rich starbursts.

Our models predict that nebular He {\sc ii} $\lambda$4686 emission
from a low-metallicity starburst should be associated with the presence 
of WC/WO stars and/or hot WN stars evolving to become WC/WO stars.
In addition, WR stars contribute to broad components beneath the 
nebular Balmer lines; the broad WR component may constitute several
percent of the total flux in the line.

We review the various techniques used to derive the WR and O star
content from integrated spectra, assess their accuracy, and propose
two new formulae to estimate the WR/O number ratio from UV or optical
spectra.

We also explore the implications of the formation of WR stars through
mass transfer in close binary systems in instantaneous bursts.  While
the formation of WR stars through Roche lobe overflow prolongs the WR
dominated phase, there are clear observational signatures which allow
the phases in which WR stars are formed predominantly through the
single or the binary star channels to be distinguished.  In particular
at low metallicities, when massive close binaries contribute
significantly to the formation of WR stars, the binary-dominated phase
is expected to occur at ages corresponding to relatively low
\hbeta\ equivalent widths.

The observational features predicted by our models allow a detailed 
quantitative determination of the massive star population in a starburst 
region (particularly in so-called ``WR galaxies'') from its integrated spectrum
and provide a means of deriving the burst properties (e.g., duration, age)
and the parameters of the initial mass function of young starbursts.
The model predictions should provide the most reliable determinations to date.
They can also be used to test the current theories of massive
star evolution and atmospheres and investigate the variation in stellar
properties with metallicity.

\end{abstract}


\twocolumn

\section{Introduction}
\label{s_intro}

Wolf-Rayet (WR) galaxies are a subset of galaxies in which broad
emission lines from WR stars are detected in the integrated galaxy spectra
(Conti 1991). (A recent review of the properties of WR stars
can be found in Maeder \& Conti 1994.)
Typically, lines of He (\heii $\lambda$ 4686), C (\ciii/\civ\ $\lambda$4650,
\civ $\lambda$5808), and N (\niii $\lambda$4640) are seen.
The strongest features are usually those at $\lambda$ 4650-4690, whose
blended emission is often referred to as the ``WR bump''. The strength 
of the various stellar emission lines has been used to estimate the
number of WR stars present in these galaxies. Typical values are in 
the range $N(WR) \sim 100 - 10^5$ (see e.g., Vacca \& Conti 1992, 
hereafter referred to as VC92).
Because WR stars are the short-lived descendants of the most massive 
stars ($M \ga 35 M_\odot$), the detection of WR emission lines in a
starburst galaxy spectrum immediately places powerful constraints on several
parameters characterizing the starburst episode: (1) the number of WR stars
relative to the number of O stars must be large, and therefore the burst
episode must have been short; (2) the initial mass function during the 
star formation episode must have extended to large masses (at least beyond
35 $M_\odot$); and (3) the time elapsed since the burst ended must be
less than a few Myrs. Thus, the presence of large numbers of WR stars
in a starburst galaxy can be used as a signpost indicating a recent
burst of massive star formation.

The first models which attempted to quantify these constraints on the WR 
populations in starbursts were introduced by Arnault, Kunth \& Schild (1989).
The comparison of their models with observations of 
$\sim$ 45 extragalactic \hii\ regions and/or galaxies with detected WR emission lines
led to several important conclusions which were largely confirmed by later studies:
(1) The observed decrease of the WR bump/\hbeta\ intensity ratio with metallicity
($Z$) and the observed variations in the WR and O star populations
in the local environment are due to the same effect, namely
the strong influence of $Z$ on stellar mass loss leads to a significant 
decrease in the WR population at low $Z$ (Maeder, Lequeux \& Azzopardi 1980); and 
(2) The derived WR/O number ratio indicates that star formation must have
occurred over only a short period compared to the lifetime of massive stars.

Synthesis models including more recent evolutionary tracks and predicting
a large variety of observational quantities were constructed by Mas-Hesse 
\& Kunth (1991a), and later updated by Cervi\~{n}o \& Mas-Hesse (1994).
The WR features included in their models are the 4650 WR bump and
\civ\ $\lambda$ 5808. Mas-Hesse \& Kunth (1991b, 1997) used these models to
analyze the spectra of nine WR galaxies.

The use of the WR bump to determine the WR population presents several difficulties.
The feature is composed of several stellar emission lines, whose strengths
can vary substantially depending on the relative numbers of the various types of
WR stars present. Furthermore, the stellar lines are often blended with nearby
nebular emission lines of Fe, He, or Ar. Kr\"uger et al.\ (1992) were the first to 
synthesize the separate \heii\ and C components of the WR bump, in an attempt to 
circumvent the problem of nebular line contamination, and to determine the dominant 
WR subtype present.
 
The effect of varying star formation rates, initial mass function,
age and metallicity on massive star populations was studied in depth
by Meynet (1995), who first stressed the importance of WR stars
of the WC sequence.
Although the extensive models of Leitherer \& Heckman (1995) include
a detailed treatment of WR and O stars, observational features due to 
WR stars are not predicted. The same also holds for the evolutionary
models of Garc\'{\i}a-Vargas, Bressan \& D\'{\i}az (1995).

Discoveries (usually serendipitous) of WR features in extragalactic 
objects have resulted from studies
covering a wide range of topics, from the primordial He abundance
determination (cf.\ Kunth \& Sargent 1983, Kunth \& Joubert 1985, 
Izotov, Thuan \& Lipovetsky 1994, 1997a; Izotov \& Thuan 1997), 
the nature of Seyfert galaxies (Heckman et al.\ 
1997), to starbursts in cooling flows (Allen 1995). A considerable
number of new observations can be expected with the new generation of 
8-10m class telescopes.
To date the total number of known WR galaxies is $\sim$ 90 
37 of which were included in the first catalogue compiled by Conti (1991).
However, quantitative analyses of the WR and
O star content has been carried out for only few ($\sim$ 20) objects for 
which high quality spectra are available (see VC92,
Kr\"uger et al.\ 1992, Mas-Hesse \& Kunth 1991b, 1997; Meynet 1995;
Schaerer 1996c; Schaerer et al.\ 1997).

In the present paper we construct new synthesis models (building upon
the models of Schaerer 1996c) 
tailored to the analysis of massive star populations in young starburst
galaxies. Using up-to-date input physics (stellar evolution tracks,
stellar atmospheres, and a new compilation of empirical WR line 
fluxes), we provide detailed predictions of many stellar and nebular
features from the UV to the optical, most of which have not been included in 
previous models.
By combining our new models with the observed strength of the WR 
emission features, as measured with medium to
high spectral resolution spectra of sufficient quality (typically
S/N $\ga$ 20), one can place constraints on the burst 
parameters which are stricter than those derived from earlier models.
With these latest models it is possible to constrain the 
IMF slope (cf.\ Schaerer 1996c), study the WC and WN populations in 
starbursts (cf.\ Meynet 1995; Schaerer \& Vacca 1996; Schaerer et 
al.\ 1997)
and to address the question of the nature of nebular \heii\ emission
in extragalactic \hii\ regions on quantitative grounds (Schaerer 1996c,
Schaerer \& Vacca 1996).
More generally the present models can be applied to
determine accurate burst parameters (age, duration, IMF etc.),
and to study massive star populations in different environments and 
their influence on the surrounding interstellar medium.

The paper is structured as follows:
The model ingredients, the input parameters, and the
synthesized quantities are described in \S \ref{s_models}.
In \S \ref{s_pops} we present predictions of the O star and
WR stars populations, including distinct WR subtypes, and discuss
the expected frequency of WR-rich starbursts.
\S \ref{s_ioniz} summarizes issues related to the evolution of the ionizing
spectrum of burst models. 
The predicted evolution of nebular H and He lines is discussed in
\S \ref{s_nebular}; the evolution of WR lines is discussed in \S
\ref{s_wrlines}.
In \S \ref{s_binary} we explore the effect of massive close binary
systems on the formation of WR stars in young bursts.
Different methods to derive the WR and O star content from integrated
UV or optical spectra are presented and re-discussed in \S \ref{s_wro}.
In \S \ref{s_discuss} we discuss applications and tests of our models
and briefly sketch future prospects for studies of massive star
populations in starbursts.

\section{Evolutionary synthesis models}
\label{s_models}
In this Section we describe the adopted model ingredients for
our evolutionary synthesis models, the most important input parameters,
and the synthesized quantities.

\subsection{Evolutionary tracks, stellar spectra, and nebular 
continuum}

As in the case of most synthesis models we assume that for most purposes 
stellar evolution can be sufficiently well described by 
evolution models for single stars. The single star assumption forms the basis
of what we call our ``standard model''. However, we also 
present calculations which include close massive binaries in an approximate 
fashion. This first order approach will allow us to investigate the uncertainties 
in the results of the standard model arising from the neglect of binaries.

\subsubsection{Single star evolution}
We use the recent evolutionary tracks of the Geneva group, which 
incorporate five different metallicities covering the 
range from $Z$=0.001 (1/20 \zsun) to $Z$=0.04 (2 \zsun) 
(see Meynet \etal 1994 and references therein). 
In particular, for massive stars we use the recent grids of Meynet \etal, which 
incorporate mass loss rates that are a factor of two higher than those adopted
in previous evolutionary models. These high-mass-loss models reproduce a large number 
of statistical observations of massive stars, including the WR to O star ratios 
in regions of constant formation (Maeder \& Meynet 1994) and they should
therefore be most appropriate for the purposes of the present work.

Previous evolutionary synthesis models which explicitly incorporated 
WR stars used different evolutionary tracks.
For example, the models of Cervi\~{n}o \& Mas-Hesse (1994) employ the Geneva tracks 
with standard mass loss rates, which yield smaller WR populations than
the preferred high-mass-loss models.
Leitherer \& Heckman (1995) adopted older tracks from 
Maeder (1991b), for which detailed comparisons with observations are shown
in Maeder (1991a).
Garc\'{\i}a-Vargas, Bressan \& D\'{\i}az (1995) use the Padua 
evolutionary tracks for which no detailed 
comparisons regarding individual WR stars and WR populations have been published.
We feel that the high-mass-loss Geneva models used in this paper should provide the 
most accurate predictions of the WR and O star populations
in a starburst region.

\subsubsection{Evolution of close massive binaries}
\label{s_bin_input}
It is fairly well established that at low metallicity the primary
mechanism for the formation of WR stars is through mass transfer 
during the evolution of massive binaries, a formation channel which should
produce primarily early-type WN stars (Maeder 1982, Maeder 1991a,
Maeder \& Meynet 1994).  Using the same evolutionary tracks as adopted
in this work, Maeder \& Meynet (1994) have shown that for regions of
constant star formation with metallicities between $\sim$ 1/10
$Z_\odot$ (SMC) and $\sim$ 1.75 $Z_\odot$ (M31), both the observed
ratio of the total number of WR stars to the number of O stars present
(the WR/O star ratio) and the subtype distribution of WR stars (WC/WR,
WNL/WR, and WNE/WR) can be well explained if a few percent of
the O stars become WR stars of mainly WNE subtype {\em as a
consequence} of Roche lobe overflow (RLOF).

To explore the implications of WR stars formed through mass transfer in
massive close binaries we proceed in the following simplified manner.
We adopt the recent evolutionary calculations from de Loore \&
Vanbeveren (1994) at different metallicities assuming an initial mass
ratio of $q=$0.6, and Case B mass transfer.  The calculations of de
Loore \&  Vanbeveren (1994) include initial primary masses from 9 to 40
\msun\ and neglect the possibility of subsequent WR formation of the
secondary. Primaries with initial masses $M_1 \ga$ 40-50 \msun\ should
in general avoid Roche lobe overflow (RLOF) (cf.\ Vanbeveren 1995);
even if they do indeed experience RLOF their evolution should be nearly
indistinguishable from that of single stars (Langer 1995). For initial
masses $M_{\rm ini} \la$ 9 \msun\ the formation of WR stars is not
expected.  Since the calculations of de Loore \&  Vanbeveren (1994) are
not available at exactly the same metallicities as the single star
models, we used the tracks with metallicities closest to those for the
single star models.  (We adopted their solar metallicity models for $Z
\ge $0.02, their LMC models for $Z=$0.008, and their SMC models for
$Z\le$0.004).

Our primary goal is to derive, as a function of the age of the burst, 
the {\em total number of WR stars} (and their subtype distribution) 
formed from mass transfer in binary systems.
For this purpose it is necessary to determine for each initial mass, whether
or not at the given burst age the primary has already experienced
RLOF leading to WR formation, and if it has, what the resulting WR subtype
is. This is done in a straightforward manner from
the lifetimes in the WR phases as given by de Loore \&  Vanbeveren 
(1994)\footnote{We note that the H-burning lifetimes 
$t_{\rm H}$ from de Loore \&  Vanbeveren (1994, Table 1) are 
larger (typically by $\sim$ 20 \%) than those from the single 
star models used here. The O star lifetimes given in Vanbeveren (1995)
agree, however, with those of standard evolutionary models
(Schaller et al.\ 1992).
To obtain a more consistent time for the onset of Case B Roche lobe 
overflow we therefore adopt the $t_{\rm H}$ lifetimes from
Meynet et al. This leads to an earlier appearance ($\sim$ 1 Myr)
of WR stars formed from mass transfer than would be obtained
by using the lifetimes from de Loore \&  Vanbeveren. The duration of 
the binary WR phases are taken directly from de Loore \&  Vanbeveren 
(Table 1).}.

One additional parameter $f$ is required to derive the total number 
of WR stars formed through the binary channel, where
$f$ is the initial fraction of stars that are primaries in close binary 
systems (described by the parameters given above) and will thus
experience RLOF during their evolution.
Because the range in initial masses covered by the binary models corresponds 
to zero-age main sequence (ZAMS) O stars, 
our definition of $f$ turns out to be essentially identical to the one given 
by Vanbeveren (1995).
In contrast 
Maeder \& Meynet (1994) use $\Psi={\rm WR_{cb}}$/O, which represents
the fraction of O stars which become WR stars through RLOF.
Note that both definitions implicitly
neglect the evolution of the secondary, which would
in principle affect the number of both O and WR stars.
In the case of a constant star formation rate $\Psi$ and $f$ are
related by 
\begin{eqnarray}
	\Psi & \approx & f \, [t({\rm WR_{cb}})/t({\rm O_{cb}})]^\prime 
							\nonumber \\
	& = & f \, \frac{\int_{M_{\rm min}}^{M_{\rm max}} 
			 t({\rm WR_{cb}})/t({\rm O_{cb}}) \Phi(M) dM }
			{\int_{M_{\rm min}}^{M_{\rm max}} \Phi(M) dM},
\end{eqnarray}
where $t({\rm WR_{cb}})$ and $t({\rm O_{cb}})$ are the lifetimes of the primary 
of initial mass $M$ in the WR phase and in the preceding O star phase, respectively.
$\Phi(M)$ is the initial mass function, and $M_{\rm min}$ and $M_{\rm max}$ are the 
minimum and maximum initial mass of primary stars leading to WR formation by RLOF.
To first order the models of Vanbeveren (1995) give mass-independent
values for $t({\rm WR_{cb}})/t({\rm O_{cb}})$ on the order of 0.2 and 0.1
for $Z$=0.02 and $Z$ $\le$ 0.01 (see his Figs.\ 2a-4a), and hence  	
	$\Psi \approx (0.2-0.1) \times f$.

For numerical comparisons with Vanbeveren (1995)
the standard mass loss tracks are the most appropriate.
As Fig.\ 12 of Maeder \& Meynet (1994) shows,
the value of $\Psi$ required to reproduce the observed WR/O number 
ratio ($\Psi \sim$ 0.08 to 0.01 for $Z \le$ 0.02) is fully compatible 
with the values of $f$ advocated by Vanbeveren (1995).
Yet another definition of the fraction of massive close binaries is
adopted in the detailed models of Vanbeveren, Van Bever \& de Donder (1997).
From their Table 1 (solar metallicity models using observational constraints)
one can, however, easily derive the values resulting for $\Psi$. One finds 
$\Psi \sim$ 0.06-0.08, as expected for standard evolutionary
models (see Maeder \& Meynet, Fig.\ 12). This indicates that the 
assumed binary frequency of Vanbeveren, Van Bever \& De Donder is in agreement 
with the work of Maeder \& Meynet (1994).
From the comparison between the theoretical and observed WR/O ratio in regions
of constant star formation using the high mass loss tracks (Fig.\ 11 of Maeder 
\& Meynet 1994) and the approximate (metallicity dependent) conversion of 
$\Psi$ to $f$ (see above) one obtains $f \sim$ 0.1 to 0.25 for $Z$=0.002 to 0.02.
For the calculations presented here we adopt $f=0.2$.

Knowing the number of WR stars formed by mass transfer, we now
proceed to derive their impact on the most important observable 
quantities, i.e.~on nebular recombination lines and on the emission
in the broad WR features. To this end one only requires a
knowledge of the ionizing flux and the number of WR stars 
in the different WR phases (cf.~\S\ \ref{s_wrinput}).
We adopt an average Lyman continuum flux of $Q_0({\rm WR}) = 10^{49} \, 
{\rm photons \, s^{-1}}$, independently of the WR subtype. 
This value roughly corresponds to the average contribution of 
(single) WR stars derived from burst models at the time 
($\sim$ 5 Myr) where binary stars are first expected to be formed
(cf.\ Fig.\ \ref{fig_meanq0_age}).
As a comparison the value used by VC92 is
$Q_0({\rm WR}) = 1.7 \times 10^{49}$. The contribution of the ionizing flux
in particular affects the H recombination lines, which are 
generally used as a reference for relative line intensity 
measurements.
The broad WR line emission is treated as for single stars
(see \S\ \ref{s_wrinput}). Since no information is available on
the WC subtype from the binary models they are assigned the WC4
subtype corresponding to the dominant subtype at low $Z$ where the
role of binaries is of largest importance.

Our procedure completely neglects the detailed
evolution of the binary stars in the H-R diagram. 
Except for the Lyman continuum, we neglect the continuum flux 
contribution of primaries in close binary systems, 
which affects only the predicted equivalent widths.
This assumption is reasonable as long as 
the UV and optical continuum fluxes are
dominated by single stars or stars in wide binary systems.
The secondaries are assumed to evolve as single
stars. The expected increase of their Lyman continuum flux after 
mass transfer (due to possible ``rejuvenation'') can be neglected 
since the primary WR star will dominate the EUV output.
We feel that this approach is a reasonable first order approximation
to study the effect of WR formation through the binary channel on 
a stellar population.
For more detailed studies the number of new parameters (mass
ratio distribution, initial binary separation and more; 
see e.g.~Dalton \& Sarazin 1995, Vanbeveren, Van Bever \& De Donder
1997) is large and several uncertainties associated with binary
evolution (mass loss during Roche lobe overflow, mixing of secondary
etc.; see e.g.~Vanbeveren, Van Bever \& De Donder 1997,
Braun \& Langer 1995) should be taken into account.
Population synthesis models which include binaries and are 
applicable to WR galaxies have also been presented by
Cervi\~{n}o \& Mas-Hesse (1996), Cervi\~{n}o, Mas-Hesse \& Kunth (1996)
and Vanbeveren, Van Bever \& De Donder (1997).
First results from our work have been reported in Schaerer \& Vacca (1996).

\subsubsection{Emergent fluxes}
To describe the spectral evolution we rely on three different sets of 
theoretical models:

{\em 1)}
For the main sequence evolution of massive stars we use the spectra
produced by the combined stellar structure and atmosphere ({\em
CoStar}) models of Schaerer \etal (1996a, b) and Schaerer \& de Koter
(1996). The models, which include \nlte\ effects, line blanketing, and
stellar winds, cover the entire parameter space of O stars.  As shown
by Schaerer (1996b) and Schaerer \& de Koter (1997) they represent
significant improvements over previous calculations, particularly for
the ionizing fluxes.  In practice we use the model set given by
Schaerer \& de Koter (1997) for stars with initial masses $M_{\rm ini}
\ge$ 20 \msun.
Additionally we restrict the use of these models to the following
\teff--\logg\ domain: $\logg \ge$ 2.2 and 
        $\logg < 5.71 \times \log \teff - 21.95$.
The latter restriction corresponds roughly to an O9/B0 spectral type.
This parameter range spans the domain of validity of the 
models and yields ionizing fluxes which merge smoothly with those
calculated from the Kurucz (1992) models described below.
For solar metallicity we consistently use the corresponding 
{\em CoStar} models. At subsolar metallicities we adopt the
low metallicity ($Z=$0.004) calculations of Schaerer \& de Koter.
For $Z=$0.04 we use a new set of {\em CoStar} models 
(Schaerer 1996, unpublished).

{\em 2)}
For stars less massive than 20 \msun we use the line-blanketed plane-parallel 
LTE atmosphere models of Kurucz (1992) calculated with a microturbulent
velocity of $v_{\rm turb}=$ 2 \kms. The Kurucz models cover a large 
range of metallicities, characterized by the iron abundance [Fe/H]. 
For each set of evolutionary tracks 
(characterized by a particular value of the total metal abundance $Z$) 
we adopted the Kurucz model atmospheres with the corresponding value of 
[Fe/H]. The assignments for the specific metallicities considered
in this paper (dictated by the metallicities of the evolutionary tracks)
were as follows: we used the Kurucz models with [Fe/H] = -1.5 for $Z = 0.001$;
[Fe/H] = -1 for $Z = 0.004$, [Fe/H] = -0.5 for $Z = 0.008$,
[Fe/H] = 0 for $Z = 0.020$, and [Fe/H] = $+0.3$ for $Z = 0.040$.
More accurate assignments are not necessary since the variations
of the stellar spectra with metallicity are much less important than 
those due to changes of the evolutionary tracks.

{\em 3)}
Stars in the WR phases are described by the spherically expanding
\nlte\ models of Schmutz, Leitherer \& Gruenwald (1992).
We identify the core radii and core temperatures of 
their models with the radius and \teff\ from the wind-free hydrostatic 
stellar model (see Schaerer 1996a for distinction and a discussion).
Test calculations using the subphotospheric radii of 
Schaerer (1996a) yield essentially identical results.
The additional parameters which control the emergent spectrum of WR 
stars are the mass loss rate \mdot\ (taken from the stellar evolution 
models) and the terminal velocity \vinf\ of the wind. 
For WN and WC stars we use the fitting formulae from Schaerer 
(1996a, Eqs.~8 and 9), 
which account for the strong dependence of \vinf\ on the WR subtype.
For WO stars we adopt \vinf=4000 \kms, the average value of 
WO3 and WO4 stars (Kingsburgh \& Barlow 1995, Kingsburgh, 
Barlow \& Storey 1995). 
It should be noted that
the atmosphere models of Schmutz, Leitherer \& Gruenwald (1992) used here
include only He and do not account for variations in the emergent 
spectra due to possible composition or metallicity effects. 
These authors argue that uncertainties in the hydrodynamic structure
should have larger effects than the neglect of metals.
For WN stars, pilot studies carried out by Schmutz (1991) indicate that
the effects of line blanketing in WR atmospheres produce moderate changes in 
the predicted spectra, although Crowther et al.\ (1997) find a significant 
reduction of the ionizing flux for a line-blanketed WNL model.
For WC/WO stars models which incorporate the abundant species of C and O 
(cf.\ Gr\"afener et al.\ 1997) have just been developed.
Once line-blanketed atmosphere models for WR stars have matured and become
widely available, they will be incorporated in our synthesis routines.

In addition to the stellar continuum, we have included the nebular 
continuum spectrum in our models. As first shown by Huchra (1977) the 
nebular contribution becomes important when hot stars providing
a large number of ionizing photons are present. To treat the
nebular emission in a simplified way we assumed the emitting gas has an 
electron temperature $T_e = 10000$ K, an electron density $N_e=100 \, 
{\rm cm^{-3}}$, and a helium abundance of 10 \% by number relative to hydrogen. 
The monochromatic luminosity of the gas, which is 
proportional to the number of Lyman continuum photons $Q_0$, is 
given by
\begin{equation}
L_\lambda = \frac{c}{\lambda^2} \frac{\gamma_{\rm total}}{\alpha_B} 
		f_\gamma Q_0,		
\label{eq_nebcont}
\end{equation}
where $\alpha_B$ is the case B recombination coefficient for 
hydrogen, and $f_\gamma$ is the fraction of ionizing photons 
absorbed by the gas. 
For the continous emission coefficient $\gamma_{\rm total}$ we used 
the atomic data from the tables of Aller (1984) and Ferland (1980)
for the wavelength regimes shortward and longward of 1 $\mu$m, 
respectively, to account for 
free-free and free-bound emission by hydrogen and neutral helium, 
as well as for the two-photon continuum of hydrogen\footnote{We
ignore two-photon continuum emission for $\lambda >$ 1 $\mu$m 
and assume $\gamma_{\rm HeI}=\gamma_{\rm H}$.}.

\subsection{Input parameters}
\label{s_input}
Evolutionary synthesis consists of calculating the properties of 
an entire stellar population as a function of time. In this work we 
concentrate on the time evolution of young populations experiencing star
formation on timescales short compared to the evolutionary 
timescales of massive stars. As an idealized case we assume an 
instantaneous burst occurring at time $t=0$.
The time evolution of the burst is followed by isochrone synthesis
based on a subroutine kindly provided by Georges Meynet.

The parameters defining the burst population are the following:
the initial metallicity $Z$, the initial mass function (IMF), and 
the total mass of stars formed in the burst. 
The last parameter serves only as a normalization constant and is 
not relevant for the quantities discussed in this work.

For the IMF we adopt a power law $\phi(m)=dN/dm \propto m^{-\alpha}$
between the upper and low cut-off masses, $M_{\rm up}$ and
$M_{\rm low}$ respectively. The IMFs in massive-star forming regions
in the Local Group generally have slopes ($\alpha$) between 2 and 2.8
(see e.g., Maeder \& Conti 1994; Massey et al.\ 1995b),
in agreement with the Salpeter (1955) value of $\alpha=$ 2.35.
We adopted the Salpeter IMF for our ``standard'' model.
For the quantities and burst ages analyzed in this work the influence 
of the lower mass cut-off is negligible, and we use $M_{\rm low}=$ 0.8 \msun.
A large number of observations indicate that the most massive stars
have masses of the order of 100 \msun or larger (e.g.\ Kudritzki 1996). 
We adopt $M_{\rm up}=$ 120 \msun\ for all calculations.
For an accurate description of the evolutionary status of stars
close to the lower mass limit for WR formation, $M_{\rm WR}$, we use the
tabulated values from Maeder \& Meynet (1994).

Results from calculations using variations of the ``standard'' input 
parameters are not presented here, but are available in electronic 
format (see Appendix).

\subsubsection{WR emission lines}
\label{s_wrinput}
To predict the broad stellar emission lines we have compiled average 
line fluxes for Of, WNE, WNL, WC, and WO stars.
Our compilation, given in Tables \ref{wn_lines} and \ref{wc_lines}, includes 
the strongest emission lines of H, He, C, and N in the UV and optical spectral 
ranges: \heii\ $\lambda$1640, the N~{\sc iii/v} $\lambda$4640 blend, 
C~{\sc iii/iv} $\lambda$4650, \heii\ $\lambda$4686, the \heii\ 
$\lambda$4861+\hbeta\ blend (denoted as 4861), the 
\heii +\civ\ $\lambda$5411 blend,
\ciii\ $\lambda$5696,  the \civ\ $\lambda\lambda$5808-5812 multiplet 
(denoted as 5808), and the \heii\ $\lambda$6562 +\halpha\ blend 
(denoted as 6560).
All other optical emission lines from WR stars are expected to be
much weaker, and therefore have not been included in our model 
calculations.
For comparison with previous models we also synthesize the two
broad ``WR bumps'' (the N~{\sc iii/v}+C~{\sc iii/iv}+\heii\ blend
at $\lambda \sim$ 4650, and \civ\ $\lambda$5808).
The calibration from Smith (1991) of these features is reported in 
Table \ref{ta_smith} for completeness.

For Of and WN stars we use the absolute flux in \heii\ $\lambda$4686
(col.\ 9 of Table \ref{wn_lines}) as a reference and compute line strength 
ratios for all other
lines relative to this reference. For WC and WO stars we use \civ\ 
$\lambda$5808 (col.\ 10 of Table \ref{wc_lines}) as the line strength reference.  
All line fluxes are expressed in terms of these ``reference lines''.

\begin{table*}
\caption{Adopted line luminosities for broad stellar emission lines from
WN stars. Mean absolute line luminosities are given for the reference 
line \heii ~4686 in ${\rm erg~ s}^{-1}$ for each subtype. Mean fluxes ratios
with respect to this reference line are given for the other lines. Below each 
mean value are the standard deviation and the number of stars included in the mean.}
\centerline{
\begin{tabular}{lclclllll}
\\ \hline \\
Subtype & & $\frac{1640}{4686}$ & $\frac{4640}{4686}$ & $\frac{4861}{4686}$ &
            $\frac{5411}{4686}$ & $\frac{5808}{4686}$ & $\frac{6560}{4686}$ & 4686 \\ 
 \\ \hline \\
%
WNE & Mean     & 7.95 & \ldots & 0.129 & 0.137 & 0.074 & 0.202 & $5.2 \times 10^{35}$ \\
    & $\sigma$ & 2.47 & \ldots & 0.031 & 0.022 & 0.027 & 0.130 & $2.7 \times 10^{35}$ \\
    & $N$      & 28   & \ldots & 7     & 38    & 5     & 7     &  26                  \\
 \\
WNL & Mean     & 7.55 & ~0.244/0.616$^a$ & 0.309 & 0.107 & 0.062 & 1.596 & $1.6 \times 10^{36}$\\
    & $\sigma$ & 3.52 & 0.151/0.519      & 0.107 & 0.022 & 0.036 & 1.511 & $1.5 \times 10^{36}$\\
    & $N$      & 17    & 5/7             & 11    & 30    & 10    & 4     &  19         \\
 \\
 \hline \\
\end{tabular}
}
$^a$ The first set of values for N {\sc iii} 4640 were derived from LMC WNL stars and were 
used for models with $Z < \zsun$; the second set of values were derived from Galactic WNL
stars and were used for models with $Z \geq \zsun$.\\
\label{wn_lines}
\end{table*}

\begin{table*}
\caption{Adopted line luminosities for broad stellar emission lines from
WC/WO stars. Mean absolute line luminosities are given for the reference line \civ ~
5808 in ${\rm erg~ s}^{-1}$ for each subtype. Mean fluxes ratios with respect 
to this reference line are given for the other lines. Below each mean
value are the standard deviation and the number of stars included in the mean.}
\centerline{
\begin{tabular}{lcllllllll}
\\ \hline \\
Subtype & & $\frac{1640}{5808}$ & $\frac{4650}{5808}$ & $\frac{4686}{5808}$ &
            $\frac{4861}{5808}$ & $\frac{5411}{5808}$ & $\frac{5696}{5808}$ & 
            $\frac{6560}{5808}$ & $5808$ \\
\\ \hline \\
WO  & Mean      & 2.65 & 0.51 & 0.386 & \ldots & 0.026 & \ldots & 0.078 & $1.1 \times 10^{36}$\\
    & $\sigma$  & 1.44 & 0.18 & 0.158 & \ldots & 0.012 & \ldots & 0.065 & $0.2 \times 10^{36}$\\
    & $N$       & 2    & 3    & 3     & \ldots & 3     & \ldots & 2     & 3                  \\ \\
WC4 & Mean      & 2.14 & 1.71 & \ldots & 0.019 & 0.023 & 0.021 & 0.045  & $3.0 \times 10^{36}$ \\
    & $\sigma$  & 1.09 & 0.53 & \ldots & 0.009 & 0.009 & 0.018 & 0.014  & $1.1 \times 10^{36}$ \\
    & $N$       & 16   & 23   & \ldots & 11    & 14    & 11    & 19     & 18                   \\ \\
WC5 & Mean      & 5.14 & 2.53 & \ldots & 0.022 & 0.036 & 0.077 & 0.074  & $9.8 \times 10^{35}$ \\
    & $\sigma$  & 4.43 & 0.49 & \ldots & 0.013 & 0.018 & 0.033 & 0.011  & $2.3 \times 10^{35}$ \\
    & $N$       & 5    & 9    & \ldots & 7     & 8     & 9     & 8      & 2                    \\ \\
WC6 & Mean      & 7.60 & 2.98 & \ldots & 0.047 & 0.054 & 0.184 & 0.112  & $8.9 \times 10^{35}$ \\
    & $\sigma$  & 6.17 & 0.52 & \ldots & 0.012 & 0.006 & 0.083 & 0.037  & $1.9 \times 10^{35}$ \\
    & $N$       & 6    & 10   & \ldots & 9     & 9     & 10    & 10     & 3                    \\ \\
WC7 & Mean      & 10.22& 3.21 & \ldots & 0.069 & 0.076 & 0.579 & 0.177  & $1.4 \times 10^{36}$ \\
    & $\sigma$  & 5.25 & 0.49 & \ldots & 0.024 & 0.027 & 0.214 & 0.028  & $0.6 \times 10^{36}$ \\
    & $N$       & 10   & 15   & \ldots & 13    & 14    & 15    & 14     & 4                    \\ \\
WC8 & Mean      & 13.69& 3.30 & \ldots & 0.096 & 0.091 & 1.968 & 0.580  & $3.5 \times 10^{35}$ \\
    & $\sigma$  & 14.58& 0.45 & \ldots & 0.037 & 0.031 & 0.649 & 0.284  & $3.1 \times 10^{35}$ \\
    & $N$       & 5    & 7    & \ldots & 5     & 6     & 7     & 8      & 3                    \\ \\
WC9 & Mean      & 4.47 & 4.46 & \ldots & 0.294 & 0.135 & 3.705 & 1.243  & $2.0 \times 10^{35}$ \\
    & $\sigma$  & 1.25 & 0.97 & \ldots & 0.095 & 0.033 & 0.514 & 0.256  & $2.0 \times 10^{35}$ \\
    & $N$       & 3    & 17   & \ldots & 17    & 17    & 18    & 16     & 2                    \\
 \\
 \hline
\\
\end{tabular}
}

\label{wc_lines}
\end{table*}

\begin{table}
\caption{Average ``WR bump'' line luminosities for WN and WC stars.
Logarithm of line luminosities in $\rm {erg s^{-1}}$.
From Smith (1991, LMC stars)}
\centerline{
\begin{tabular}{lrr}
\\ \hline \\
	& C~{\sc iii/iv}4650+\heii 4686 & \civ 5808 \\
\\ \hline \\
WN7	& 36.5  & 35.1 \\
WC4	& 36.7  & 36.5 \\
\\ \hline
\end{tabular}
}
\label{ta_smith}
\end{table}

{\em Of stars:}
The strongest emission lines in Of stars are \heii\ 4686 and the N~{\sc iii/v}
4640 blend. The \heii/\niii\ ratio shows considerable variations
between individual OIf stars. We adopted a ratio of 
$L(4640)/L(4686)=0.1$, which should strictly be considered a lower limit.
For the \heii\ 4686 line we adopted an average luminosity   
determined from one O3If/WN6 and one O6Iaf 
star in the LMC (Crowther 1996, private communication). This yields a
value of $2.5 \times 10^{35}$ erg s$^{-1}$.
This is a factor 2.6 lower than the line flux from the four 
O3f/WN stars in R136a recently analyzed by de Koter, Heap \& Hubeny 
(1997); for those stars $L(4686)=6.5 \times 10^{35}$ erg s$^{-1}$
(de Koter 1996, private communication).
As can be seen the line luminosity of the R136a stars is compatible 
with the average value derived for WNE stars.

{\em WN stars:}
Unfortunately, there is no single reference that provides intrinsic
fluxes for all the lines in WN spectra that we wish to include in our
models. We compiled flux measurements from the studies of Morris (1995)
and Smith, Shara, \& Moffat (1996, hereafter SSM96).  Although Morris
(1995) and SSM96 present results for many of the same stars, the two
sets of measurements have only two lines in common: \heii\ $\lambda
4686$ and  \heii\ $\lambda 5411$. We used the 5411/4686 line ratio and
the absolute value of the 4686 flux to check that the two sets of
measurements were consistent.

We began by dereddening the observed fluxes for the Galactic stars in
both samples with the color excess values given by Morris et
al.\ (1993), with the revisions given by Morris (1995).  For three
stars for which values of $E(B-V)$ were not given we adopted the values
given by Hamann et al.\ (1993).  We used the reddening curve given by
Seaton (1979) and $R_V=3.2$.  For the LMC stars we used the extinction
curve given by Seaton (1979) for the Galactic foreground reddening and
the average LMC extinction curve derived by Vacca (1992; see also
Morris et al.\ 1993) with $R_V = 3.2$ for the reddening
within the LMC. We assumed a maximum foreground reddening value of
$E(B-V)_{fg} = 0.03$ mag.

We adopted the subtype classifications given by SSM96. This rarely
resulted in a change of more than one subtype with respect to older
classifications (e.g., Conti \& Massey 1989). We then divided the
sample into WNE and WNL subclasses. The WNE subclass consists of stars
with subtypes between WN2 and WN4, while the WNL subclass consists of
stars with subtypes between WN6 and WN9. This division agrees with the
definitions and assignations given by Conti \& Massey (1989).
Consistent subclass identifications could not be assigned to the WN5
subtype and these stars were excluded from the averages. We then
computed the average dereddened 5411/4686 ratios for LMC and Galactic
stars for the two subclasses in each sample. It was found that the averages 
for the Morris sample and those for the SSM96 sample could be brought into
agreement only if stars with low hydrogen abundances (WNb and WNo stars, 
as classified by SSM96) were excluded from the WNL averages. WR 43 was 
the only exception to this rule of excluding objects classified by SSM96 as 
having low H abundance from the WNL averages; although SSM96 classifiy this
object as a WN6o, Drissen et al.\ (1995; see also Dessart \& Crowther 1997) 
find that it is composed of three
WN6h stars. Correspondingly, we excluded those stars containing a substantial 
amount of hydrogen (WNh or WN(h) stars, again as classified by SSM96) from the WNE 
averages. Once the mean line ratios were found to be consistent between the 
two samples, we combined the samples. The flux ratios of
\heii\ $\lambda$1640/\heii\ $\lambda$4686 and
\heii\ $\lambda$5411/\heii\ $\lambda$4686 are taken from Morris (1995),
while the other line strength ratios are taken from SSM96. 

As shown by SSM96 the strength of the N $\lambda$4640 line depends on the 
initial abundance of the progenitor star. For solar metallicities and larger we
adopt the Galactic average, while for $Z<\zsun$ we adopt the lower LMC
value. For all of the other line ratios, we derived average values by
combining the LMC and Galactic stellar samples, after confirming that
the means of the two samples for each line ratio were in agreement.
We excluded from the averages any values which were found to be more
than 3$\sigma$ away from the mean; the averages were then re-computed 
with the smaller sample. This process resulted in the exclusion of
only a few individual line flux values from the calculations of the
various means.

The absolute luminosity of \heii\ for each subclass was derived from
the Morris (1995) sample of LMC stars and those Galactic stars that are
definite or probable members of cluster or associations with reliable
distance estimates. We assumed a distance of 50.1 kpc (DM=18.50) for the
LMC; distances for stars in Galactic clusters or associations were
taken from Lundstr\"om \& Stenholm (1984), with revisions given by
Moffat (1983; for NGC 3603), Garmany \& Stencel (1992), Smith, Meynet, 
\& Mermilliod (1994), and Massey, Johnson, \& DeGioia-Eastwood (1995a) 
and references therein. For WR 43/NGC 3603, we divided the total dereddened
flux measured by Morris (1993) into three components, with relative 
contributions in agreement with the line measurements of Drissen (1997, private
communication). The resulting mean WN line luminosities and line ratios, as 
well as the standard deviations about the mean and the number of stars 
included in each mean, are presented in Table (\ref{wn_lines}).

\begin{figure}[htp]
\centerline{\psfig{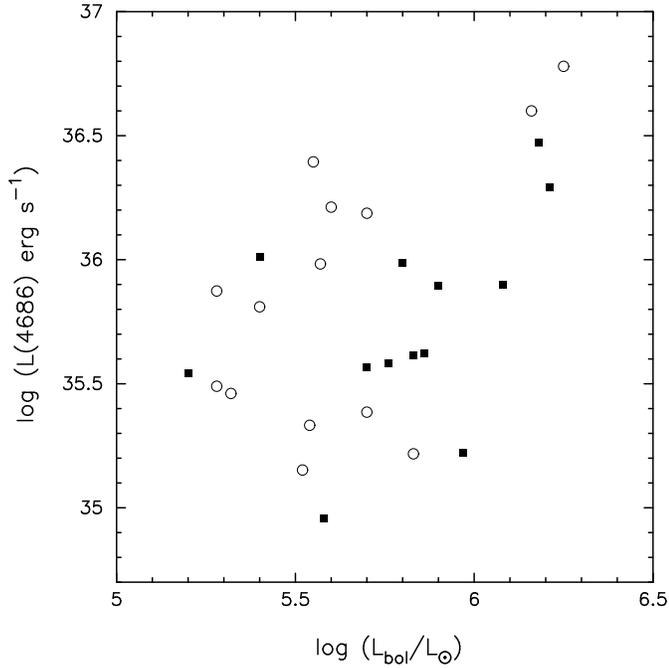}}
\caption{Luminosity in the \heii\ 4686 line vs. the bolometric luminosity
for WNL stars, from the measurements and analysis by Crowther \& Dessart 
(1997). Open circles denote LMC stars; filled squares are Galactic stars. 
Note the huge variation in line luminosity across the WNL subclass.
\label{fig_wnplot} }
\end{figure}

A comparison of the line fluxes determined for individual stars by
Morris (1995) with those measured by Crowther (1997, private
communication) and Crowther \& Dessart (1997) on substantially higher
resolution spectra reveals a systematic difference of $\sim 20-30$ \%
(in the sense that the line fluxes from Morris 1995 are larger than those of
Crowther \& Dessart 1997). It is not completely clear what the cause of
this discrepancy might be, but if we assume that the higher resolution spectra
yield more accurate line fluxes, then we conclude that our adopted line luminosities may
be $\sim 30$ \% too large. In addition, the absolute luminosity of
\heii\ $\lambda$4686 exhibits substantial variations within the WNL
subclass, and a mean luminosity for this line is rather uncertain.  The
situation is illustrated in Fig.\ \ref{fig_wnplot}, in which we plot the
line luminosities from Crowther (1997) and Crowther \& Dessart (1997) against the
bolometric luminosities derived by these authors from their spectral
analyses.  The large variation in line luminosity within the WNL
subclass is easily seen. The figure also reveals that there may be a
difference in the line luminosity between stars with $L < 10^6 L_\odot$
and those with $L > 10^6 L_\odot$. The lower luminosity stars (in the
combined Galactic plus LMC sample) have a mean \heii\ 4686 line
luminosity of $5.6 \times 10^{35}$ erg s$^{-1}$, while the higher
luminosity stars have a mean line luminosity of $3.1 \times10^{36}$ erg
s$^{-1}$.  However, the uncertainties in the mean values for these two
luminosity classes are also very large, and we have chosen to use a
constant line luminosity for all bolometric luminosities of WNL stars
in our models. The entire WNL sample from Crowther (1997) 
has a mean line luminosity of $1.1
\times 10^{36}$ erg s$^{-1}$, which agrees well with our adopted value
after the systematic differences between the Crowther (1997) and Morris
(1995) line flux measurements are taken into account.

Despite the uncertainties in the mean luminosity of the \heii\ 4686 line, our 
estimate of $1.6 \times 10^{36}$ erg s$^{-1}$ is in excellent agreement with 
the value of $1.7 \times 10^{36}$ erg s$^{-1}$ determined by Vacca (1992; 
see also VC92). Vacca (1992) demonstrated that this value, when combined with 
spatially-integrated line flux measurements, yields an estimate for the number 
of WNL stars within the 30 Doradus complex that is in good agreement with 
the number derived by directly counting objects with spectroscopic identifications.
Furthermore, Terlevich et al. (1996) used this mean luminosity value to 
determine the number of WN stars in NGC 604 from their spatially-integrated
spectra. They found that their estimate was again in excellent agreement with 
the number of WR stars found by Drissen, Moffat, \& Shara (1993) using
narrow-band {\em HST} images. In addition, Crowther \& Dessart (1997) find
that this value of the mean line luminosity yields a number for the WNL stars
within the NGC 3603 complex that is in good agreement with the observed number 
(Drissen et al.\ 1995). Unfortunately, additional tests of our value are currently not available.

To reduce the uncertainties of the WR line strengths used in the
present models it would be highly desirable to understand the observed
large variations of the line luminosities (Fig.\ 1) and their
dependence on stellar parameters. If correlations of the line strengths
with physical parameters can be established, these can be used in
future models to provide a more accurate picture of the WR populations.
It is also necessary to clarify the relationship between  WR
classifications and stellar models (cf.\ Smith \& Maeder, 1997).  For
example, the WNE and WNL classifications we have adopted are defined
according to purely phenomenological criteria and do not necessarily
correspond directly to the definitions employed in the stellar
evolutionary models. Our additional separation of the stellar sample
into objects with and without hydrogen represents a compromise between
the observational definitions of WNE and WNL stars and those employed
by the evolutionary models. With these criteria we have attempted to
link the observations and models as closely as possible. The adopted
approach should yield average line luminosities and ratios that are
representative of the properties of the classes of hydrogen-free and
hydrogen-rich WN stars.

{\em WC and WO stars:}
In view of the substantial variations in the frequency distributions of WC/WO
subtypes with environment (cf.~Smith \& Maeder 1991), as well as the strong
variations in spectral signatures with subtype, we chose to treat the various WC
subtypes separately. The assignment of the WC/WO subtype from the
evolutionary model is discussed below (\S \ref{s_synth}).

The task of compiling dereddened line fluxes was substantially easier
in the case of WC and WO stars.  Average line fluxes of WC stars have
been derived by Smith, Shara, \& Moffat (1990a, b, hereafter SSM90a,
SSM90b) and recently by Brownsberger (1995), who provided an extensive
list of dereddened line fluxes for nearly every known WC star in the
Galaxy and the LMC.  Their results agree within their quoted
uncertainties.  We adopted the dereddened line fluxes presented by
Brownsberger (1995) and computed average values of the line flux ratios
(relative to the strength of \civ\ $\lambda$5808) as a function of WC
subtype; the subtypes were assigned according to the classification
scheme of SSM90a and SSM90b.  (The method used by Brownsberger to
compute the reddening for each star and deredden the line fluxes is
identical to the method we adopted for WN stars; see above.) The
average luminosity in \civ\ $\lambda$5808 for each subtype was
determined from WC stars in the LMC and those Galactic stars located in
clusters or OB associations for which reliable distances could be
determined. Distances were taken from Lundstr\"om \& Stenholm (1984),
with revisions given by Garmany \& Stencel (1992), Smith, Meynet, \&
Mermilliod (1994), and Massey, Johnson, \& DeGioia-Eastwood (1995a) and
references therein. Unfortunately it is not possible to check for
consistency in the \civ\ line luminosity between LMC and Galactic WC
stars; the LMC contains only one WC star with a subtype later than WC4,
and there are no WC4 stars in the Galaxy with well determined
distances.  For the WC4 subtype we adopted the average
\civ\ $\lambda$5808 luminosity of the LMC stars. This value is 0.5 dex
larger than that found for WC5-7 stars in the Galaxy, which agrees with
the findings of SSM90b.  
The mean line ratios were derived by combining the LMC
and Galactic WC4 samples, after confirming that their individual mean
values were completely consistent with one another.  For all other
types, we adopted the mean line luminosity and line ratios derived from
the Galactic stars. The average line luminosities and ratios, as well
as the standard deviations about the means and the number of stars
included in the estimation of the means, are given for each subtype in
Table \ref{wc_lines}. The averages were computed iteratively, as described above 
for the WN stars, by excluding those individual values that were 
found to be more than 3 $\sigma$ from the means. This resulted in
the exclusion of only a few line flux values from the calculations
of the various means.

The emission of WC stars in the 4650 bump requires an additional
comment.  The broad \ciii/\civ\ $\lambda$4650 blend also includes the
contribution from \heii\ $\lambda$4686, which usually can not be
separated. From SSM90b (their Table 2) we find that for WC5-9 stars
\heii\ 4686 contributes on the average $\sim$ 12 \% to the 4650 blend;
SSM90a estimate \heii\ 4686 contributes 8 to 30 \% of the blend for
WC4. We include the strength of the entire 4650 blend in our models and
do not treat the \heii\ 4686 emission from WC stars separately.

For WO stars we have derived average line luminosities from the dereddened
line flux measurements of three WO3 and WO4 stars given by Kingsburgh \& 
Barlow (1995) and Kingsburgh, Barlow \& Storey (1995).  While the line
luminosities of these objects (all located in low metallicity
metallicity environments: LMC, SMC, IC1613) exhibit relatively little
scatter, they are systematically stronger than the line luminosities
for the two known galactic WO1 and WO2 stars. This difference may be
attributed to differences in their evolution (Kingsburgh, Barlow \&
Storey, 1995). Since WO stars are expected to occur mainly at low
metallicities (cf.\ \S\ \ref{s_wcwo}) it seems justified to adopt the
properties of only WO3/WO4 stars.
We used the fractional line contributions provided by Kingsburgh,
Barlow \& Storey (1995) and Kingsburgh \& Barlow (1995) to decompose
the total flux in the emission blend at 4650 \AA\ into the flux from
\civ\ $\lambda 4658$ and that from the component due to \civ\ $\lambda
4685$, and \heii\ $\lambda$ 4686. (The latter two lines would not
usually be separable in the spectra of W-R galaxies.)
From test calculations we find that other emission lines which are
prominent only in the rare WO stars are predicted to be very weak and
are therefore not included in our models.
For completeness we list their average line luminosities, which are:
$L($\ion{O}{4} $\lambda$3400$)=1.9 \times 10^{36}$ erg s$^{-1}$, 
$L($\ion{O}{6} $\lambda$3811$)=1.5 \times 10^{36}$ erg s$^{-1}$, and
$L($\ion{O}{5} $\lambda$5590$)=1.5 \times 10^{35}$ erg s$^{-1}$.
Note that we have adopted a distance of 58.9 kpc for the SMC and
661 kpc for IC 1613.

\subsection{Synthesized quantities}
\label{s_synth}
The most important synthesized quantities can be classified in 
three groups: population statistics, nebular quantities, and
WR emission lines.

\subsubsection{Population statistics}
For comparisons with star counts it is useful to derive the
numbers of stars with different spectral types.
Of particular interest for young bursts is the number of
O stars, the total number of WR stars and their distribution
among the WNL, WNE, and WC/WO subclasses.
WR stars and the various WR subclasses are defined by their surface 
abundances and their 
effective temperature from the evolutionary models (see e.g.~Maeder 
1991b). Hydrogen-free WN models are considered to be WNEs, while those
containing hydrogen are treated as WNLs.
We also estimate the number of stars belonging to the various WC subtypes 
(WC4 - WC9) and WO subclasses (cf.~below); the frequency distribution 
among these subtypes depends strongly on metallicity
(e.g.~Smith \& Maeder, 1991). We assign WC and WO subtypes
according to their surface abundance, i.e. by the (C+O)/He number ratio.
\footnote{WC/WO subtypes are assigned as follows:
WC9: (C+O)/He $\le$ 0.08, 
WC8: (C+O)/He $\in$ (0.08,0.15],
WC7: (C+O)/He $\in$ (0.15,0.25],
WC6: (C+O)/He $\in$ (0.25,0.35],
WC5: (C+O)/He $\in$ (0.35,0.43],
WC4: (C+O)/He $\in$ (0.43,1.],
WO: (C+O)/He $>$ 1.
Here the values of Smith \& Maeder (1991) have been adjusted slightly
to account for the recent C, O, and He determinations in WC4 stars by
Gr\"afener et al.\ (1997).  The correspondence between spectral types
and C, O abundance is not completely agreed upon, however (cf.
e.g.\ Eenens \& Williams 1992 and Koesterke \& Hamann 1995).
More detailed analyses,  covering a large range of WC subtypes
and including oxygen abundance determinations, will be necessary 
to assess the validity of the correlation of (C+O)/He with
WC subtype.}
Note that in some cases the WR phase can be entered 
during the core H-burning phase (cf.~Meynet \etal 1994).
We define O-type stars to be those H-burning objects with 
effective temperatures $\teff >$ 33000 K. 

We also estimate the number of Of stars and include the contribution
of their stellar emission lines to the observed emission features.
We assumed that all O stars with
surface gravities below the average value found for OIa stars
can be counted as Of stars. We used the recent 
calibration of Vacca \etal (1996) to derive $\logg$ as a 
function of \teff\ for the O Ia luminosity class. For consistency
we adopt the calibration given for the gravity $g_{\rm evol}$ 
determined from evolutionary models
(following Vacca \etal 1996).

\subsubsection{Nebular quantities}
We synthesize the following H and He nebular recombination lines
under the assumption of Case B recombination:
\halpha, \hbeta, \hei\ $\lambda$ 4471, and \heii\ $\lambda$ 4686.
The adopted electron temperature and density are the same as given
above. The emissivities and recombination coefficients are taken
from Osterbrock (1989).
The luminosity emitted in the emission lines is calculated from 
the number of Lyman, \hei, and \heii\ photons ($Q_0$, $Q_1$, and
$Q_2$ respectively) using the relations

\begin{eqnarray}
L(\halpha) & = & 1.36 \times 10^{-12} \, f_\gamma \, Q_0  
		\label{eq_q0_halpha} \\
L(\hbeta)  & = & 4.76 \times 10^{-13} \, f_\gamma \, Q_0  
		\label{eq_q0_hbeta_1} \\
L(4471)    & = & 2.22 \times 10^{-13} \, f_\gamma \, Q_1  
		\label{eq_q1_hei} \\
L(4686)    & = & 1.02 \times 10^{-12} \, f_\gamma \, Q_2  
		\label{eq_q2_heii}
\label{eq_neb_lines}
\end{eqnarray}
where $L$ is given in $[{\rm erg \, s^{-1}}]$ and $Q_i$ in 
$[{\rm s^{-1}}]$. 
$f_\gamma$ is the fraction of ionizing photons absorbed by the 
gas. A fraction $(1-f_\gamma)$ is assumed to escape the system.
Our ``standard'' model assumes an ionization bounded nebula with
$f_\gamma=1$. 
It should be noted that the strength of \hei\ $\lambda$ 4471
predicted from Eq.\ \ref{eq_neb_lines} above will be incorrect
when there are enough photons above 24.6 eV
to fully ionize helium in the nebula; this situation corresponds 
approximately to equivalent effective temperatures of $\teff >$ 40000 K.

\subsubsection{WR emission lines}
Several observational signatures of WR stars in the integrated 
spectrum of a starburst region are computed according to the prescriptions 
given in Sect.\ \ref{s_wrinput}.

\begin{table*}
\caption{Maximum WR/O ratios}
\centerline{
\begin{tabular}{lrrrrrrrrrrrrl}
\\ \hline \\
$Z$	& \multicolumn{3}{c}{WNL/O} & & \multicolumn{3}{c}{WR/O} & \\
 	& $\alpha=1.$ & $\alpha=2.$ & $\alpha=2.35$   & & $\alpha=1.$ 
			& $\alpha=2.$ & $\alpha=2.35$ & ~~~~~~ \\
\\ \hline \\
0.040	&  1.50 & 0.54  & 0.36	& & 1.50 & 0.54  & 0.36 & ($t =$ 3 Myr) \\
0.020	&  0.51 & 0.17  & 0.11	& & 0.97 & 0.32  & 0.22 & ($t \le$ 4 Myr) \\
0.008	&  0.23 & 0.12  & 0.10	& & 0.40 & 0.14  & 0.11 \\
0.004	&  0.13 & 0.058 & 0.044 & & 0.29 & 0.091 & 0.059 \\
0.001   &  0.12 & 0.033 & 0.020 & & 0.15 & 0.041 & 0.025 \\  
\\ \hline
\end{tabular}
}
\label{ta_wro}
\end{table*}
\section{Wolf-Rayet and O star populations in young starburst}
\label{s_pops}
We now discuss the model predictions regarding the massive star
populations.  We present the time evolution of the WR and O star
populations for an instantaneous burst at different metallicities. We
explicitly consider the various WR subtypes (primarily WNL and WC/WO)
and estimate the fraction of O stars which exhibit emission lines (Of stars).
The results presented here are essential for  understanding the
behaviour of the synthesized spectral lines and other observable
features which will be discussed in \S\ \ref{s_nebular} and
\ref{s_wrlines}.

\begin{figure}[htp]
\centerline{\psfig{figure=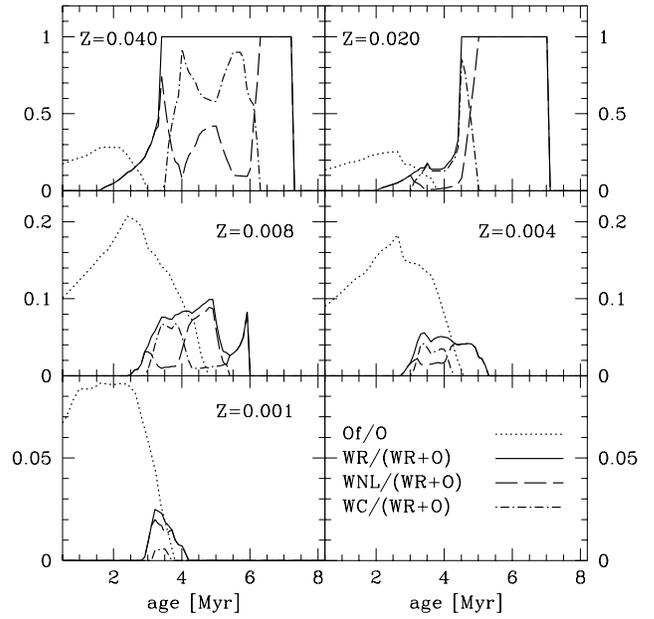,height=8.8cm}}
\caption{Predicted Of/O and WR/O star number ratios as a function of age for all
metallicities. All models are calculated for an instantaneous burst
with a Salpeter IMF. Shown are the total WR/(WR+O) (solid), the
WNL(WR+O) (dashed), the WC/(WR+O) (dashed-dotted), and the Of/O 
number ratios. Note the change of the ordinate between the top, 
middle, and lower panels.
\label{fig_wrratios} }
\end{figure}

In Figure \ref{fig_wrratios} we show the time evolution of the stellar
number ratios WR/(WR+O), WNL/(WR+O), and WC/(WR+O), as well as the
Of/O ratio for the five metallicities between $Z$=0.001 and 0.040. 
(Note that here WC stands for WC and WO stars, irrespective of the distinction 
made between these types in the calculation of the line strengths.)
The values were calculated
for an instantaneous burst (occurring at $t=0$) and a Salpeter IMF.
Variations in the relative WR and O star populations with other
parameters (IMF slope, upper mass limit, star formation rate and
different evolutionary tracks) have been discussed in detail by Meynet
(1995), and therefore will not be repeated here. Nevertheless, 
we must comment on a small discrepancy between our results for WR 
populations and those presented by Meynet
(1995). The differences in the two sets of results can be seen by
comparing, for example, our predictions for the WC population at 
$Z$=0.004 (Fig.~\ref{fig_wrratios}) with those presented in Fig.~3 from Meynet (1995).  
As Fig.~\ref{fig_wrratios} shows, we predict a strong decrease of the WC
population after $\sim$ 4 Myr, while in models of Meynet (1995) the WC
population remains dominant until the end of the WR-rich phase ($t
\sim$ 5.2 Myr).  The origin of this discrepancy resides in the use of
different techniques to interpolate between the two (discrete)
evolutionary tracks above and below the mass limit of WR formation (in
the case of $Z$=0.004, between the 60 and 40 \msun\ track). This discrepancy
occurs therefore only for ages larger than the lifetime of the least
massive star whose initial mass is larger the WR mass limit (60
\msun\ in the present case).  While Meynet (1995) performs a linear
interpolation on the lifetimes in the various WR phases (WNL, WNE, and
WC) as a function of initial mass
between those two tracks, our classification of WR subtype is based on an
interpolation of the surface abundances on the isochrone.  Meynet's
method predicts the existence of WC stars until the end of the WR
phase, while in our case models close to the WR mass limit have surface
abundances typical of the less evolved WNL stars, since mass loss is
not sufficient to expose more processed matter.  This numerical
difference explains the shorter duration of the WC phase (and hence
correspondingly longer WNL phase) in all our models compared to the
results of Meynet (1995). For the same reason the duration of the WNE
phase in our models is significantly shorter.  Although both the exact
behaviour of stars in this mass range and the value of $M_{\rm WR}$ are
somewhat uncertain, we feel that the present approach yields a more
physical description of the transition between the least massive WR
stars and the lower mass stars.

The following important points are illustrated in Figure 
\ref{fig_wrratios}:
\begin{itemize}
\item The fraction of Of/O stars (classified by the criterion given
in \S \ref{s_synth}) is typically of the order of 0.1 to 0.3 for burst
ages up to 2-3 Myr. The predicted changes with $Z$ are a result of the 
differences in the main sequence as a function of $Z$.
Due to the blueward shift of both the zero-age main sequence and the 
terminal-age main sequence (cf.\ Meynet 
et al.\ 1994), the fraction of its main sequence-lifetime a star with an given 
initial mass spends as an O star increases with lower $Z$.
This leads to the general decrease of Of/O to low $Z$.
\item The {\em maximum of the WR/O ratio} (and therefore also
the WNL/O and WC/O ratios) and the {\em duration of the WR-rich phase}
decrease rapidly with metallicity (cf.~Maeder \& Meynet, 1994; Meynet
1995).  The maximum predicted values are given in Table \ref{ta_wro}.
Note that for $Z < \zsun$ our WNL/O values are slightly larger than
those given by Meynet (1995, his Fig.~6) although both results are
based on the same evolutionary tracks. This difference is essentially 
due to the overestimate by Meynet of the WC population, as discussed above.
For $Z \ge 0.020$, the WR/O star number ratio formally becomes infinite 
at ages $t \ge$ 4.5 Myr because the O star population (defined to 
be those objects with $\teff \ge$ 33000 K) disappears completely while WR
stars are still present. Therefore, for these metallicities in Table \ref{ta_wro} 
we give values for the WR/O ratio at specific times $t < 4.5$ Myr (cf. 
Meynet 1995). 

\item In an instantaneous burst the WR-rich period (WR/O $>$ 0) is 
characterized by three distinct phases:
{\em 1)} The {\bf early WNL phase} at the beginning of the WR-rich phase
occurs before the most massive stars evolve to WC stars.
{\em 2)} The {\bf WC-rich or WC-dominated phase} with coexisting WNL stars, 
where the precise mixture of different WR subtypes depends critically on 
age and metallicity.
{\em 3)} The {\bf final WNL phase} with WR stars which have evolved from objects
with initial masses close to the lower mass limit of WR formation $M_{\rm WR}$. 
Phases occurring before and after the WR-rich period have been discussed
by Meynet (1995), to which we refer the reader for more details.

\end{itemize}

\begin{table}
\caption{Statistics of \protect\hii\ and WR galaxies. The metallicity is given in
col.\ 1. Cols.\ 2-6 indicate the predicted fraction of \protect\hii\ regions
containing: only O stars (col.\ 2), WR and O stars (3), WR but no O stars (4).
The fraction of \protect\hii\ regions dominated (in number)
by WNL stars, and WC stars is given in Cols.\ 5 and 6.
All numbers are relative fractions of the total number of predicted 
\protect\hii\ regions defined as clusters harbouring O and/or WR stars.
Col.\ 7 indicates the fraction of WR-rich \protect\hii\ regions dominated
by WC stars (in number).
}
\centerline{
\begin{tabular}{lrrrrrrrrrrrrl}
\\ \hline \\
Z	& O$_{\rm only}$ & WR+O & WR$_{\rm only}$ & WNL & WC & 
		$\rm{\frac{WC}{total WR}}$  \\
\\ \hline \\
0.040	& 0.20 & 0.26 & 0.54 & 0.48 & 0.32 & 0.40 \\
0.020	& 0.28 & 0.36 & 0.36 & 0.50 & 0.22 & 0.31 \\
0.008	& 0.40 & 0.60 &      & 0.43 & 0.17 & 0.28 \\
0.004	& 0.60 & 0.40 &      & 0.27 & 0.13 & 0.33 \\
0.001   & 0.82 & 0.18 &      & 0.18 \\  
\\ \hline
\end{tabular}
}
\label{ta_stats}
\end{table}

\subsection{Frequencies of WR-rich starbursts}
An interesting result from the work of Meynet (1995) is the predicted
frequency of WR-rich objects in young starbursts. In Table \ref{ta_stats}
we present the revised frequencies predicted from our models.
For an instantaneous burst scenario one can easily determine
the time during which the population contains only O stars
(``O$_{\rm only}$''), both WR and O stars (``WR+O''), or
WR stars and no O stars (i.e.~B instead of O stars: ``WR$_{\rm only}$'').
The values given in Table \ref{ta_stats} (cols.~2-4) are the fraction 
of time spent in these phases with respect to the total time when O 
and/or WR stars are present.
Columns 5 and 6 give the fraction of time during which the WR population
is expected to be dominated (in number) by stars of WNL and WC subtypes
respectively. The last column (col.~7) gives the fraction of 
WR-rich starbursts that are expected to be dominated by WC stars (i.e.
the values in column 7 are derived by dividing the values in column 6 by
the sum of the values in columns 3 and 4).
Note that the values in cols.~5-7 differ from those of Meynet (1995, 
Table 1, high mass loss models) whose results overestimate the 
relative duration of the WC and WNE phases with respect to the WNL phase 
as discussed above\footnote{In Table 1 of Meynet (1995), columns 1 and 2 for the 
high mass loss models at $Z$=0.004 are misprints. All other values in 
common with Meynet are in agreement.}.

For an instantaneous burst of star formation, the phases during which 
O and/or WR stars are present in the population should lead to a
strong emission line spectrum.
Therefore, the fractions given in Table \ref{ta_stats} 
should represent the fraction of \hii\ regions (extragalactic \hii\ 
regions or alike objects) which should be dominated by the various
stellar types.
Although comparisons with observations are difficult due to problems 
regarding the statistical completeness of a given sample, estimates 
derived from low metallicity samples yield encouraging results 
regarding both the frequency of the WR phase and the occurrence
of WC stars (Schaerer \& Vacca 1996). 
Systematic studies of the WR and O content in statistically 
complete samples should be possible in the near future.

\subsection{WC/WO populations in starbursts}
\label{s_wcwo}
From both observational and theoretical studies it is well known that
the subtype distribution of WC stars changes
dramatically with metallicity (cf.~Smith \& Maeder, 1991).  In
particular, late type WC stars are found only in high metallicity
environments (e.g., Conti \& Vacca 1991; Philips \& Conti 1992), while
all WC/WO stars are found to be WC4 or WO types in low metallicity
regions.

\begin{figure}[htp]
\centerline{\psfig{figure=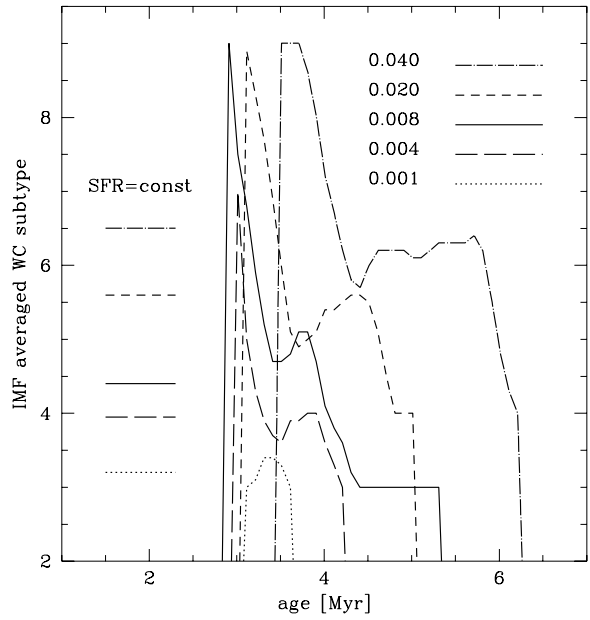,height=8.8cm}}
\caption{Temporal evolution of the IMF-averaged WC subtype for 
all metallicities ($Z$=0.040: dashed-dotted, 0.020: short dashed,
0.008: solid, 0.004: long dashed, 0.001: dotted). WO stars are designated
by a WC3 subtype
\label{fig_wcsubtype} }
\end{figure}

WC stars are expected to evolve towards earlier subtypes during their
lifetime, an evolution that reflects the change in surface abundance as
larger amounts of processed matter are brought to their surface (Maeder
1991b).  Therefore in a burst at a given metallicity, the distribution
of WC/WO subtypes is expected to evolve with time.  To quantify this
behaviour we have calculated the IMF-weighted mean WC subtype
(henceforth called IMF-averaged subtype), which reflects the average WC
subtype of a mixed population of WC stars.  We assigned WO stars to the
WC3 subtype.  The time evolution of the IMF-averaged WC subtype at
different metallicities is plotted in Figure \ref{fig_wcsubtype}.  The
horizontal lines on the left indicate the average value for regions of
constant star formation.

Figure \ref{fig_wcsubtype} nicely illustrates the qualitative changes
of the WC population not only with metallicity but also with time.  As
expected this figure shows the predominance of WO and early type WC
stars at low metallicity. Similarly, at a given $Z$ the average WC
subtype decreases with time due to the increase of the surface (C+O)/He
abundance ratio along each evolutionary track.  Late WC stars are
therefore expected only at early times during the WC-rich phase
(typically at $t \sim$ 3-4 Myr), or during later times in a high
metallicity region.


\section{Evolution of the ionizing spectrum}
\label{s_ioniz}
In this Section we discuss the behaviour of quantities
directly related to the ionizing flux of the stellar population.

\subsection{Ionizing fluxes}
\begin{figure}[htp]
\centerline{\psfig{figure=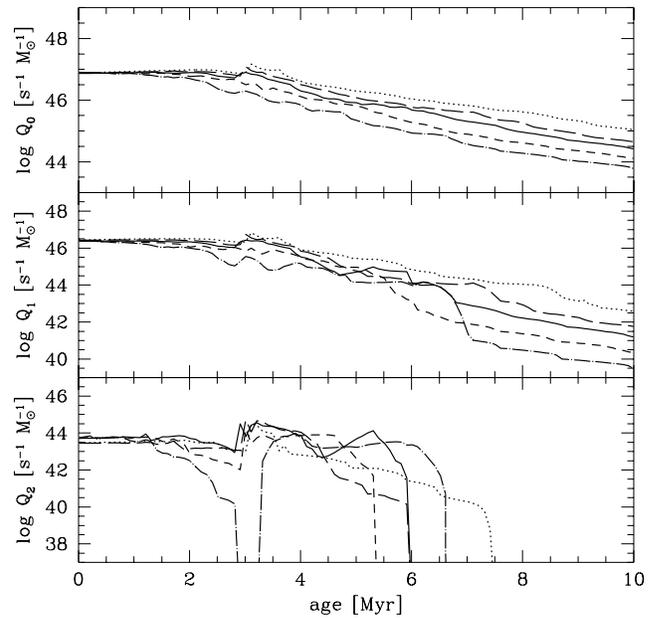,height=8.8cm}}
\caption{Temporal evolution of the ionizing photon flux of a stellar
population (normalised to a total mass of 1 \msun) for all metallicities.
{\em Top panel:} Lyman continuum flux,
{\em Middle:} Flux in the He~{\sc i} continuum,
{\em Lower panel:} flux in the He~{\sc ii} continuum.
Same symbols as in Fig.~\protect\ref{fig_wcsubtype}
\label{fig_qi_time} }
\end{figure}

Figure \ref{fig_qi_time} presents the predicted time evolution of the
photon luminosity in the ionizing continua of hydrogen
($Q_0$), \hei\ ($Q_1$), and \heii\ ($Q_2$). Note that all values
are normalised to a population of a total mass of 1 \msun\ assuming
a Salpeter IMF from 0.8 to 120 \msun.
However, the qualitative properties discussed below are applicable
to most IMFs.

The metallicity dependence of the ionizing continuum of an integrated population 
is determined primarily by variations in the evolutionary tracks; 
changes due to blanketing effects in the atmosphere models are 
relatively minor (e.g.\ Cervi\~{n}o \& Mas-Hesse 1994).
The behaviour of the Lyman continuum flux is controlled essentially by the
main sequence evolution. In WR-rich phases the maximum contribution
to the Lyman continuum provided by these stars is typically $\sim$
30 \% (cf.~also below).
Main sequence stars also govern the behaviour of $Q_1$, although
the picture is somewhat more complicated by the fact that WR stars 
contribute a non-negligible fraction of the \hei\ ionizing flux.
In very young bursts ($<$ 3 Myr) the \heii\ ionizing continuum is 
dominated by O stars.
The small differences with metallicity in this case are due to changes 
in the stellar wind properties, which dominate the opacity in the 
\heii\ continuum (see Schaerer \& de Koter, 1997).
During the WR-rich phases $Q_2$ is clearly dominated by the contribution
of WR stars. The time and metallicity dependence of $Q_2$ is therefore
directly related to the duration of this phase and the total number of
WR stars present (the ``strength'' of the WR phase 
illustrated in Fig.~\ref{fig_wrratios}).

\begin{figure}[htp]
\centerline{\psfig{figure=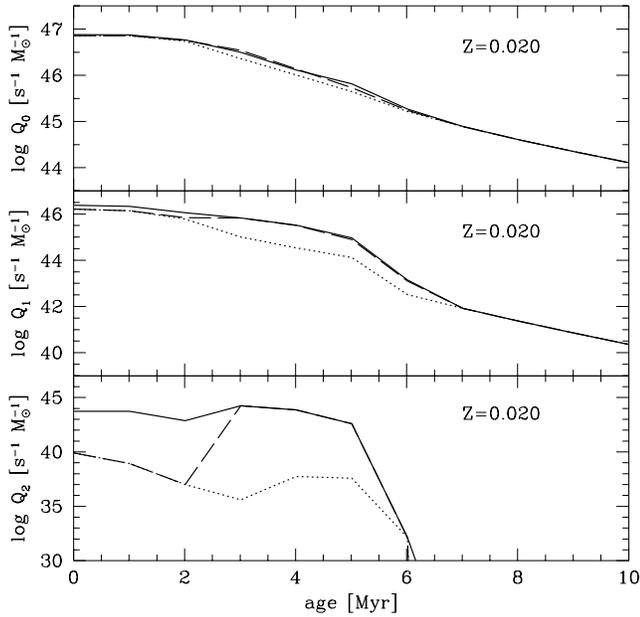,height=8.8cm}}
\caption{Comparison of predicted ionizing fluxes using different atmosphere
models. Shown is the standard model for $Z$=0.020. 
Dotted: using plane parallel, LTE, line blanketed models of Kurucz (1992) only;
Dashed: same as dotted but using spherically expanding \nlte\ models 
for WR stars (Schmutz, Leitherer \& Gruenwald, 1992);
Solid: same as dashed but using spherically expanding, line blanketed, \nlte\
{\em CoStar} models for O stars. The top, middle, and bottom panel show 
$Q_0$, $Q_1$,and $Q_2$ respectively
\label{fig_qi_models} }
\end{figure}

\begin{figure}[htp]
\centerline{\psfig{figure=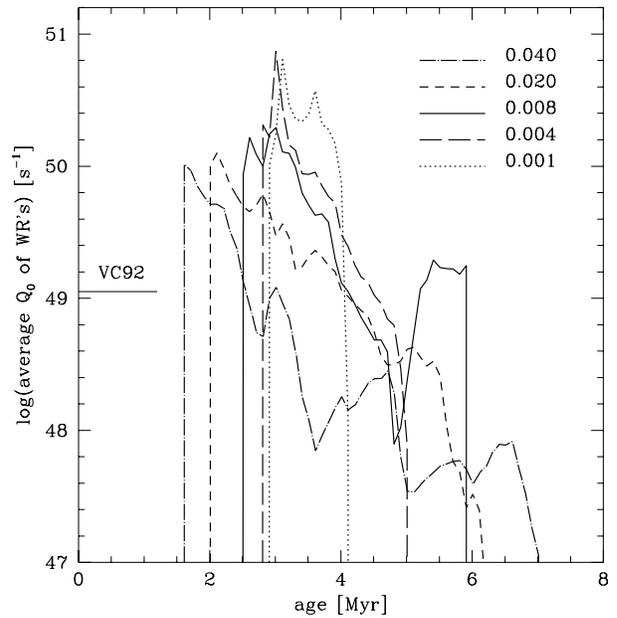,height=8.8cm}}
\caption{Time evolution of the average Lyman continuum luminosity provided
per WR star in instantaneous burst models.
The thin solid horizontal line indicates the value of $Q_O^{\rm WNL}$ from
Vacca \& Conti (1992).
Same symbols as in Fig.~\protect\ref{fig_wcsubtype}
\label{fig_meanq0_age} }
\end{figure}

What implications do the recent stellar atmosphere models have for the
predictions of the ionizing flux of a stellar population?  As
discussed in detail by Schaerer \& de Koter (1997) the inclusion of
\nlte\ effects, line blanketing, and stellar winds in the {\it CoStar}
models leads to significant changes in the ionizing fluxes of O stars.
The predictions of these models appear to be supported by observations
of \hii\ regions (Stasi\'nska \& Schaerer, 1997).  Figure
\ref{fig_qi_models} illustrates the different predictions for the
ionizing fluxes for an instantaneous burst at solar metallicity from the
various atmosphere models for O and WR stars. The solid line shows our
standard model, using the {\em CoStar} models for O stars and the
fluxes from Schmutz, Leitherer \& Gruenwald (1992) for WR stars.  The
dashed line was derived using the Kurucz (1992) plane parallel, LTE
model atmospheres models instead of the {\em CoStar} models for O stars.
Adopting Kurucz model atmospheres for both O and WR stars (using the
wind corrected effective temperature from the evolutionary models)
yields the dotted line.

Figure \ref{fig_qi_models} (cf.~dotted and dashed) illustrates the
importance of using appropriate atmosphere models for WR stars to
describe the ionizing fluxes in WR-rich phases (between $\sim$ 3 and 6
Myr).  The role of WR stars in this context is discussed in detail by
Schmutz, Leitherer \& Gruenwald and Garc\'{\i}a-Vargas (1996).  As can
be seen from Fig.~\ref{fig_qi_models} (comparing dashed and solid
lines) the inclusion of {\em CoStar} models for O stars leads to the
following changes with respect to 
synthesis models relying on older (and less complete) stellar 
atmospheres:
{\em 1)} In young bursts $Q_2$ increases by at least 4 orders
of magnitudes. The absolute value of the ionizing flux in the 
\heii\ continuum can therefore reach values similar to those attained 
during the WR-rich phase.
{\em 2)} In the first few million years of a burst the \hei\
ionizing flux increases by typically 60 \%.
{\em 3)} Small differences in the Lyman continuum flux are  
due mostly to larger values of $Q_0$ for supergiants with $M = 20$ to 40 \msun.

\subsubsection{The ionizing flux of WR stars in starbursts}
As mentioned above the contribution of WR stars to the ionizing
spectrum of a stellar population is not negligible, especially 
at the shortest wavelengths.
In this context, it is interesting to examine the average Lyman
continuum luminosity emitted per WR star, $Q_0({\rm WR})/N({\rm WR})$, 
where $Q_0({\rm WR)}$ is the total Lyman luminosity emitted from all
WR stars and $N({\rm WR})$ is the number of WR stars present.
This quantity is plotted in Fig.~\ref{fig_meanq0_age} where we show
the temporal evolution of $Q_0({\rm WR})/N({\rm WR})$ for all 
metallicities.

For each metallicity the average $Q_0$ emitted per WR star generally
decreases with time as expected from the decrease of their average 
luminosity with progressing age.
The increase of the maximum average Lyman luminosity (and also of the
time-averaged  $Q_0({\rm WR})/N({\rm WR})$) with decreasing $Z$ simply 
reflects the higher average luminosity of WR stars at low $Z$, or 
equivalently, the larger value of $M_{\rm WR}$ - the minimum mass of 
WR formation - at lower $Z$ (cf.\ \S\ \ref{s_input}).
It is interesting to note that the average value of $Q_0$ per WR
star shows variations over typically more than one order of magnitude
during the WR rich phase. As will be discussed below (\S\ \ref{s_nwro})
this variation affects estimates of WR/O ratios derived from \heii\ 
$\lambda$4686 and \hbeta\ line strengths; 
in the past such estimates were calculated assuming a single dominant 
WR subtype and a constant value of $Q_0$ per WR star
(e.g.~$Q_0=1.7 \times 10^{49}$ s$^{-1}$ for WNL stars, VC92).

\section{Evolution of nebular lines of hydrogen and helium}
\label{s_nebular}
\subsection{\protect\hbeta}
Hydrogen recombination lines, particularly \hbeta\, serve as a
fundamental reference for measurements of other emission lines. In
addition, under the assumption of an ionization bounded nebula, their
luminosities provide an estimate of the ionizing flux present and their
equivalent widths can be used as an accurate indicator of the age of
the population in the case of an instantaneous burst of star formation
(a coeval population), as first pointed out by Copetti, Pastoriza \&
Dottori (1986).  A critical discussion of this and other age indicators
for young starbursts can be found in Stasi\'nska \& Leitherer
(1996).

\begin{figure*}[htp]
\centerline{\psfig{figure=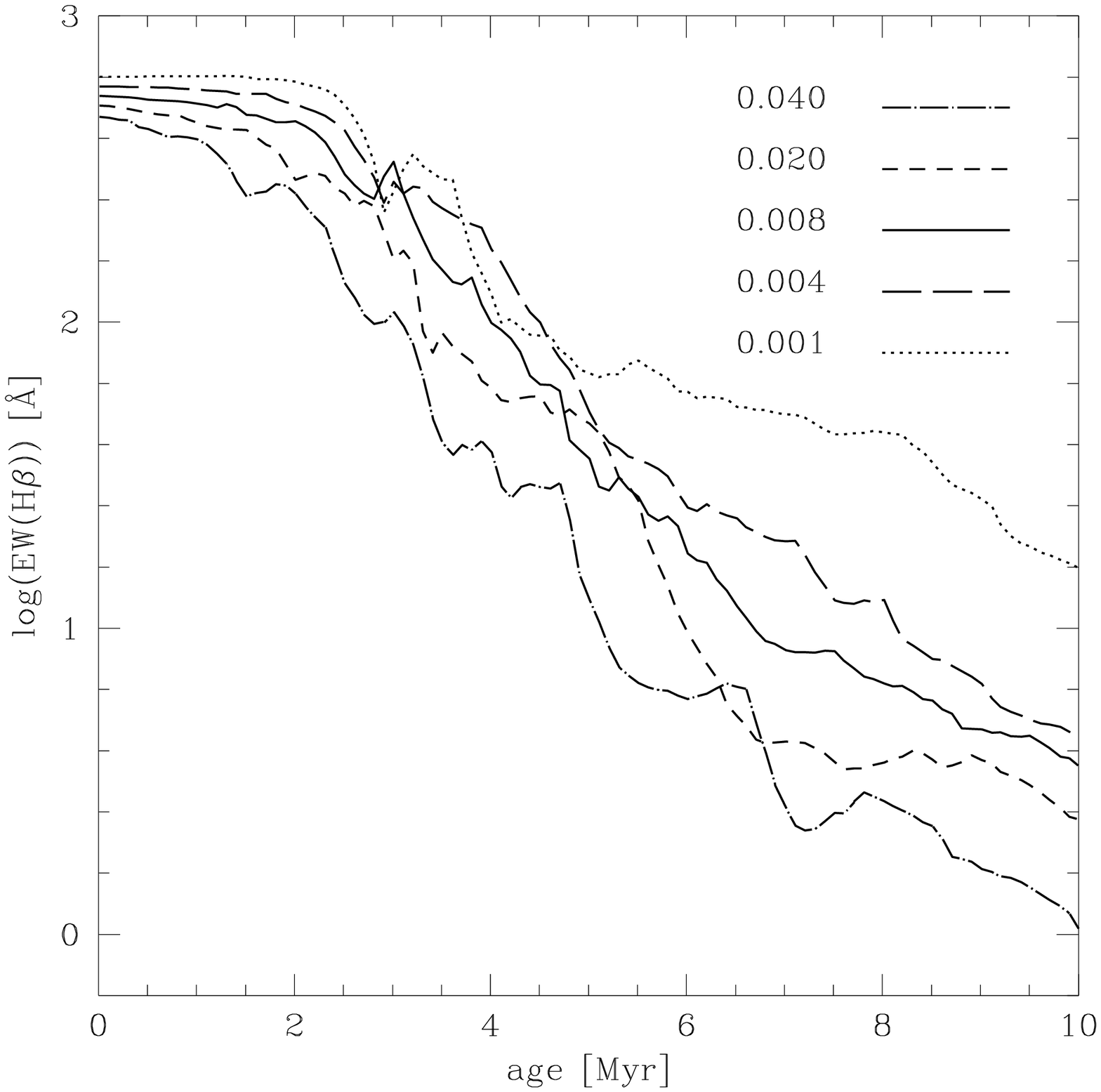,height=8.8cm}
	    \psfig{figure=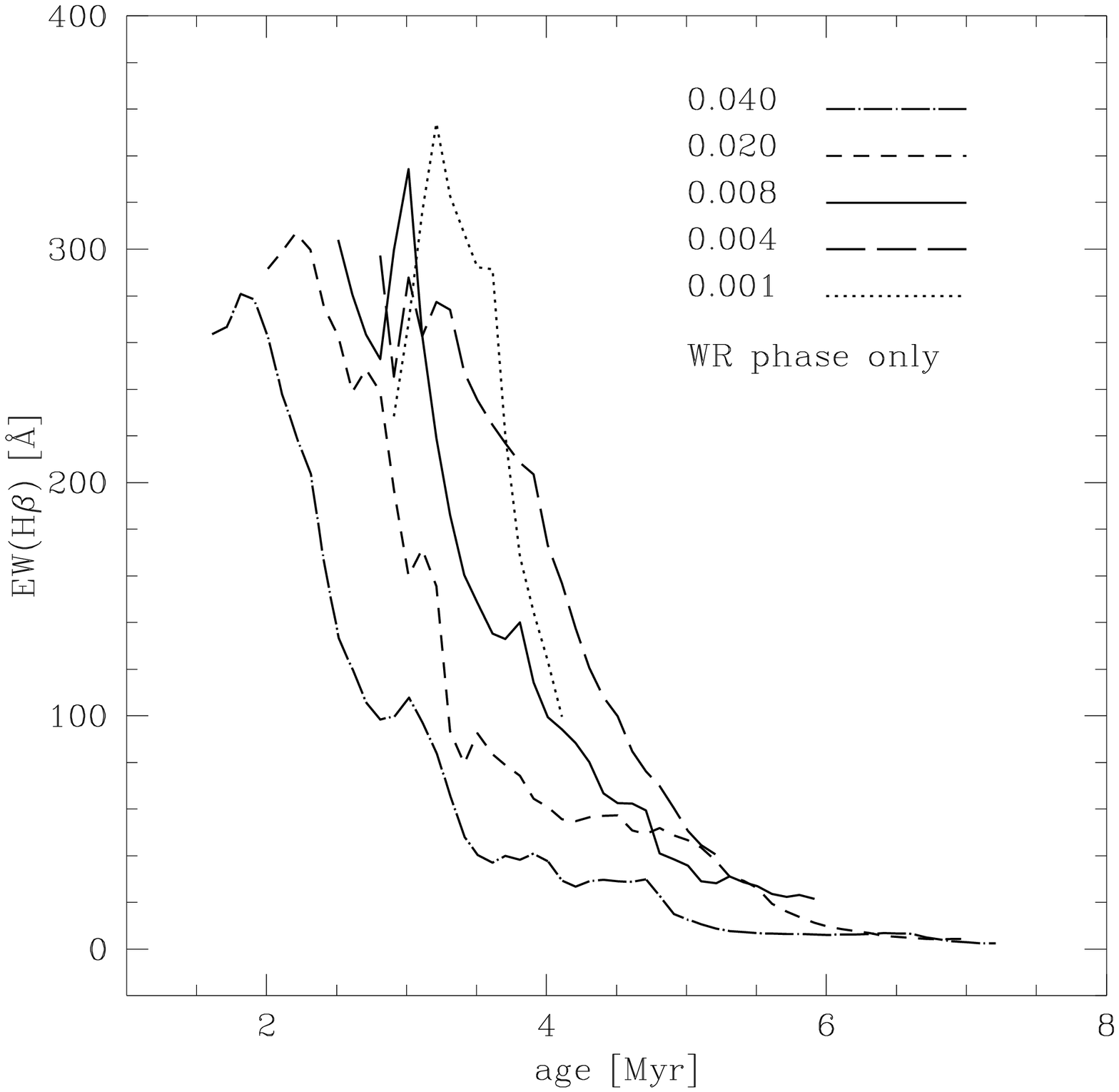,height=8.8cm}}
\caption{Time evolution of the \protect\hbeta\ equivalent width for an 
instantaneous burst at the given metallicity.
Same symbols as in Fig.~\protect\ref{fig_wcsubtype}.
{\em Left panel:} Logarithm of $W({\rm H}\beta)$ for ages from 0 to 10 Myr.
{\em Right panel:} Evolution of $W({\rm H}\beta)$ during the WR rich phases
of the burst. Horizontal lines on the margins show the maximum/minimum 
$W({\rm H}\beta)$ during the WR phase
\label{fig_hbeta_age} }
\end{figure*}

The temporal evolution of the \hbeta\ equivalent width $W(\hbeta)$
for all metallicities is shown in Fig.\ \ref{fig_hbeta_age} (left panel).
The right panel shows $W(\hbeta)$ only during WR rich phases.
The results presented in these figures can be combined with 
results presented below (cf.\ Figs.~\ref{fig_lhb_opt}, \ref{fig_lhb_opt_2})
to determine the WR and O star populations in a region from observational
quantities.

\subsection{\protect\heii\ $\lambda$4686}
\begin{figure}[htp]
\centerline{\psfig{figure=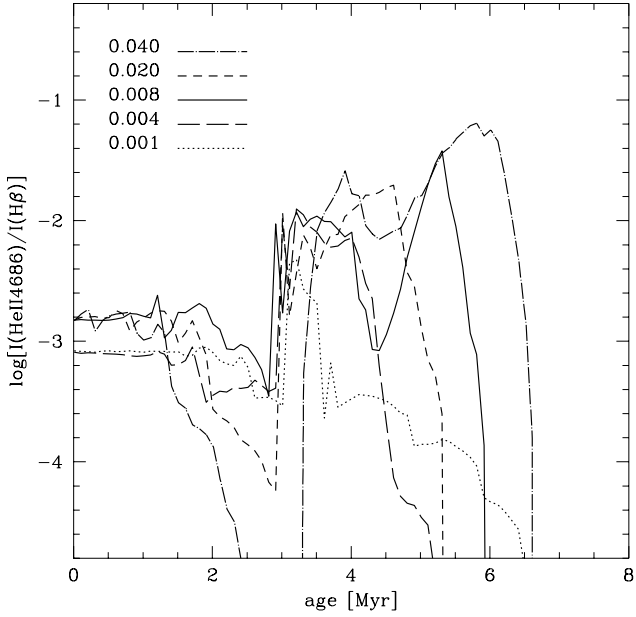,height=8.8cm}}
\caption{Predicted {\em nebular} emission line ratio of 
\protect\heii\ $\lambda$4686/\protect\hbeta\ as a function
of age.
Same symbols as in Fig.~\protect\ref{fig_wcsubtype}.
See Figs.\ \protect\ref{fig_ews_opt} and \protect\ref{fig_heii_lhb}
for the total WR and nebular 4686 emission
\label{fig_heii_whb} }
\end{figure}

Figure \ref{fig_heii_whb} shows the predicted {\it nebular} \heii\ 
$\lambda$4686 emission for all metallicities; this figure was presented
in Schaerer (1996c). Here we briefly reiterate the most important points
demonstrated in this figure.

For young bursts dominated by O stars ($t \la$ 3 Myr) typical values of
$I($\heii$)/I(\hbeta)$ are between $5 \times 10^{-4}$ and $2 \times 10^{-3}$.
As mentioned above (cf.~\S\ \ref{s_ioniz}), due to the strong
\heii\ continuum flux obtained from the most recent and physically
complete O star models, these values are $\sim$ 4 order of magnitudes
larger than predictions from other current synthesis models.  
However, the $I($\heii$)/I(\hbeta)$ values are still 
too low to be detectable in nebulae around O stars or in young \hii\ regions.

\begin{figure}[htp]
\centerline{\psfig{figure=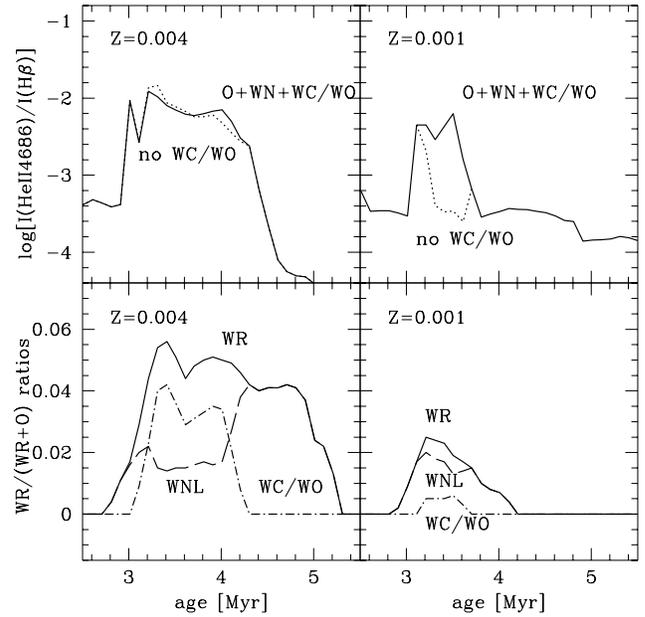,height=8.8cm}}
\caption{
{\em Lower panels:} total WR/(WR+O) (solid lines), WNL/(WR+O) (dashed),
and WC/(WR+O) (dashed-dotted) number ratios as a function of time
for $Z$=0.004 (left) and $Z$=0.001 (right). 
{\em Upper panels:} Nebular \protect\heii\ 
$\lambda$4686/\protect\hbeta\ as a function of age for $Z$=0.004
(left) and $Z$=0.001 (right).
The dotted lines show the predicted \protect\heii\ 
$\lambda$4686/\protect\hbeta\ when WC stars are excluded.
This shows that both WC and (hot) WN stars contribute to the nebular
\protect\heii\ emission
\label{fig_heii_wc} }
\end{figure}

During the subsequent WR rich phase $I($\heii$)$ / I$(\hbeta)$ values of
10$^{-2}$ to 10$^{-1}$ are obtained.
In order to determine which sources contribute to the high value of
$I($\heii$)/I(\hbeta)$ we have plotted the 
time evolution of 4686/\hbeta\ during the WR phase for two selected
metallicities in Fig.~\ref{fig_heii_wc} (upper left panel:
$Z$=0.004, upper right: $Z$=0.001) 
and the corresponding WR/(WR+O), WNL/(WR+O), and WC/(WR+O) number 
ratios (lower left panel: $Z$=0.004, lower right: $Z$=0.001).
%
%
This Figure indicates that large 
nebular \heii/\hbeta\ values are obtained during the 
WC/WO phase as well as during a short time before and after the WC phase
(cf.\ Fig.~\ref{fig_wrratios} and Schaerer 1996c).
The high excitation is provided by WC/WO stars and WN stars that 
are about to evolve to the former phase.
In addition, the relative contributions of WN and WC/WO stars are a strong
function of both age and metallicity.  This is illustrated  by presenting
the value of $I($\heii$)/I(\hbeta)$ obtained when 
the contribution of WC/WO stars to the ionizing flux is excluded.
For $Z$=0.004 almost no change in the line intensity ratio is seen because
the nebular \heii\ emission is due primarily to hot WN
stars with low hydrogen surface abundances (H mass fraction $<$ 0.1).
For $Z$=0.001 the high $I($\heii$)/I(\hbeta)$ ratio is due mainly to 
highly evolved WC/WO stars
(in this case primarily the WO subtypes, cf.~Fig.~\ref{fig_wcsubtype},
which have a large $Q_2/Q_0$) despite their small numbers relative to
WN stars.  In conclusion, both WC/WO and hot WN stars produce the nebular 
\heii\ emission.  Their relative contributions, however, depend very sensitively
on metallicity and (especially) age.

The measurement of the observed flux in a nebular \heii\ $\lambda$ 4686
line from a starburst region is complicated by the possible presence of a broad
stellar component. In some cases, the narrow nebular component will be
superposed on a broad feature (cf.\ e.g.\ VC92). However, in many instances,
the broad feature may completely dominate the emission, which makes the
identification and measurement of a nebular component nearly impossible
with low to medium resolution spectroscopy. The interplay between the
two components will be discussed in \S\ \ref{s_4686}.
 
\section{Evolution of broad stellar emission lines}
\label{s_wrlines}
In this Section we discuss the behaviour of the broad 
emission lines included in our calculations (see \S\ \ref{s_wrinput}).
Two different types of illustrations are useful for comparisons with
observational data:
{\em 1)} Plots of the equivalent widths of a feature as a function
	of age, and
{\em 2)} Line intensity ratios with respect to \hbeta\ as a function
	of the \hbeta\ equivalent width, which serves as an age indicator.
Depending on the spectroscopic observations 
(type of object, spatial resolution, quality of observations etc.), and 
several other factors one or the other approach may be favoured.
Figures of the first type depend only on stellar quantities,
since in most cases nebular continuum emission is fairly small.
Equivalent widths may, however, be reduced by an underlying population.
Intensity ratios of the WR lines with respect to \hbeta, for example,
have the  advantage of being relatively 
independent of such dilution effects.
On the other hand comparisons of this type rely on the 
assumption of an ionization bounded nebula, and may be affected by 
the relative spatial distribution of the stellar and nebular
emission, and the extent of the emission compared to the size of the 
aperture used for the spectroscopic observations.

\subsection{UV lines: \protect\heii\ $\lambda$1640}
\begin{figure}[htp]
\centerline{\psfig{figure=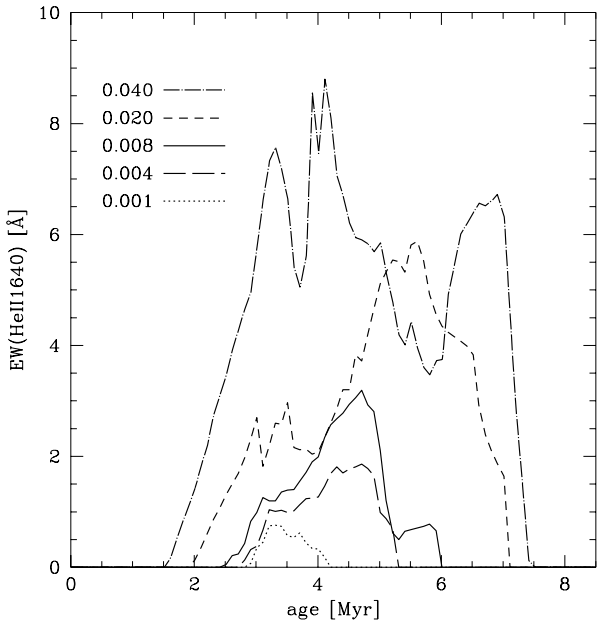,height=8.8cm}}
\caption{Evolution of the \protect\heii\ $\lambda$1640 equivalent
width.
Same symbols as in Fig.~\protect\ref{fig_wcsubtype}
\label{fig_ews_uv} }
\end{figure}

Figure \ref{fig_ews_uv} shows the temporal evolution of the
\heii\ $\lambda$1640 equivalent width for all metallicities.  As with
all quantities related to WR stars, the substantial variation with
metallicity is due to the strong dependence of WR evolution on $Z$
(cf.~Maeder \& Meynet 1994).  For subsolar metallicities $W(1640)$ is
always predicted to be smaller than $\sim$ 4 \AA.  For these
metallicities, the time dependence of $W(1640)$ follows the total
WR/(WR+O) ratio quite closely, as discussed in \S\ \ref{s_1640}.  Since
for $Z >$ 0.02 WN stars dominate the 1640 emission (late WC stars have
a lower 1640 emission), 
stronger variations are found between the WC and WN dominated phases.

\subsection{Optical lines}
\label{s_opt}
\subsubsection{The traditional WR bumps at low spectral resolution}
\label{s_wrbumps}
\begin{figure}[htp]
\centerline{\psfig{figure=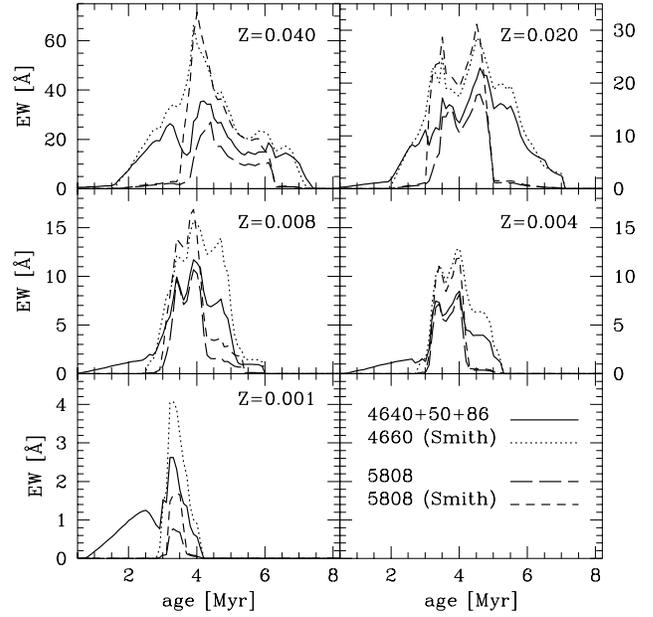,height=8.8cm}}
\caption{Evolution of WR bump equivalent widths for all metallicities.
Solid lines: 4660 bump including N~{\sc iii/v} $\lambda$4640,
C~{\sc iii/iv} $\lambda$4650, and \protect\heii\ $\lambda$4686.
Long-dashed: \protect\civ\ $\lambda$5808.
Both following the prescriptions in Tables \protect\ref{wn_lines}
and \protect\ref{wc_lines}.
Dotted and short-dashed lines show the WR bump equivalent widths
using the prescriptions of Smith (1991; cf.\ Table 
\protect\ref{ta_smith})  
\label{fig_wrbumps} }
\end{figure}

In Figure \ref{fig_wrbumps} we plot the evolution of the equivalent
width of the two WR emission blends prominent in the optical wavelength
regime; these are the strongest WR features found in the spectra of WR
galaxies.  Adopting the line fluxes from Tables \ref{wn_lines} 
and \ref{wc_lines} we have
calculated the 4660 \AA\ bump (solid line) from the sum of N~{\sc
iii/v} $\lambda$ 4640,  C~{\sc iii/iv} $\lambda$ 4650, and
\heii\ $\lambda$ 4686; the 5808 \AA\ emission blend (long-dashed) is
due to the \civ\ multiplet.  The 4660 bump may be contaminated
by additional nebular emission from \heii, \hei, and forbidden Fe and
Ar lines observation at medium or low resolution, the distinction between
the individual components of the bump is often difficult.  This makes
comparisons to the total ``bump'' necessary in many cases, although
working with the individual WR lines would be preferred.

As Fig.~\ref{fig_wrbumps} shows, the equivalent width of the 4660 bump
is usually expected to be larger than that of 5808.  In WC rich phases,
however, $W(5808)$ can reach values comparable to that of 4660.  Note that the
appearance of the 4660 bump before the WR phase ($0.5 \la t \la 2.5$
Myr) is due to stellar \heii\ $\lambda$4686 emission from OIf stars
(cf.~below).  Compared to the recent models of Cervi\~no \& Mas-Hesse
(1994) \footnote{Their adopted WR bump luminosities are $L_{4660}=2 \times
10^{36}$ erg s$^{-1}$ for WN stars and $L_{4650}=6 \times 10^{36}$ erg
s$^{-1}$ for WCs, which differ by 
$\sim$ 20-15 \% 
from our values for WNL and WC4 stars.}our predicted equivalent widths 
for the 4660 bump are larger by up to
a factor of $\sim$ 3 for low $Z$. The larger values are essentially due
to the high mass loss evolutionary tracks adopted in the present work,
which lead to larger WR/O star ratios in better agreement with
observations (see Maeder \& Meynet 1994).

For comparison we also plot the equivalent widths obtained using
the bump luminosities of Smith (1991) given in Table \ref{ta_smith}
(dotted and short-dashed line in Fig.~\ref{fig_wrbumps}).
In this case larger EWs for the two WR bumps are expected
as can be seen from a comparison of Tables \ref{wn_lines} and
\ref{wc_lines} with \ref{ta_smith}.
Indeed the 4660 bump luminosity from Smith (1991), derived from 
\heii\ $\lambda$4686 of 5 LMC WN6-7 stars, represents
an upper limit for all LMC WN stars (cf.~Vacca 1992), even
if the N~{\sc iii/v} $\lambda$ 4640 contribution is included.
It must, however, also be reminded that N~{\sc iii/v} $\lambda$ 4640
seems to be metallicity dependent (cf.\ \S\ \ref{s_wrinput}).
We feel that, for most cases, our procedure of allowing
for the strong variations of WC subtypes and including more recent
calibrations for WN stars should yield more reliable predictions
for the WR bumps than calculations based on the data of Smith (1991).

\begin{figure}[htp]
\centerline{\psfig{figure=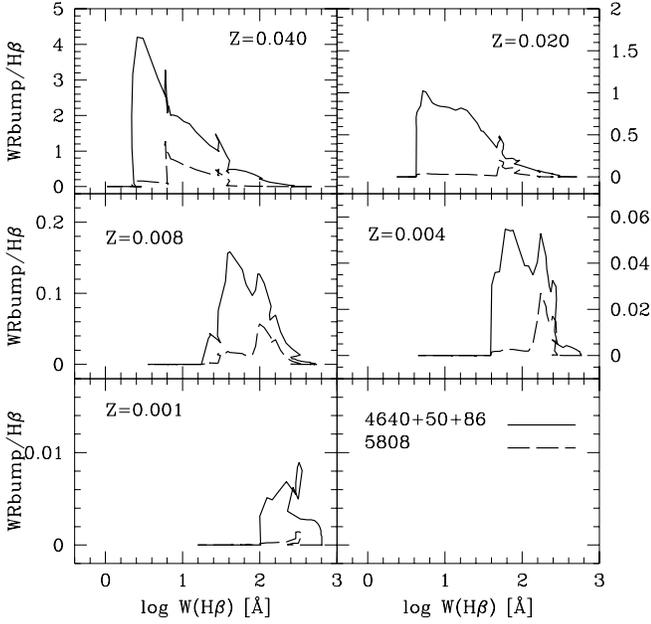,height=8.8cm}}
\caption{Evolution of WR bump intensities relative to \protect\hbeta\ 
for all metallicities.
Solid: 4660 bump, 
long-dashed: \protect\civ\ $\lambda$5808, computed as in Fig.\ 
\protect\ref{fig_wrbumps}. The less accurate values using the prescription
of Smith (1991) is not shown here.
Certain cases show non-uniqueness of line intensities when plotted 
against $W(\protect\hbeta)$ (e.g.\ for $Z$=0.001, but cf.\ also other
Figs.). 
This behaviour is due to
the slight increase of the \protect\hbeta\ with time at the beginning
of the WR phase (see Fig.\ \protect\ref{fig_hbeta_age}), which is in turn
related to the contribution of WR stars to the Lyman continuum flux
\label{fig_wrbumps_hb} }
\end{figure}

Figure \ref{fig_wrbumps_hb} shows the relative WRbump/\hbeta\ line
intensities of the two bumps. Given the steepness of the spectrum the 
maximum intensity ratio 5808/\hbeta\ is considerably smaller than 
4660/\hbeta\ in constrast to the maximum of the equivalent widths 
(cf.\ above).
The strong increase of the peak 4660/\hbeta\ with metallicity is 
in large part due to the much more rapid decrease of \hbeta\ for
large metallicities (cf.\ Arnault, Kunth \& Schild 1989).

\subsubsection{WR bumps at medium and high spectral resolution}
\label{s_wropt}
\begin{figure*}[htp]
\centerline{\psfig{figure=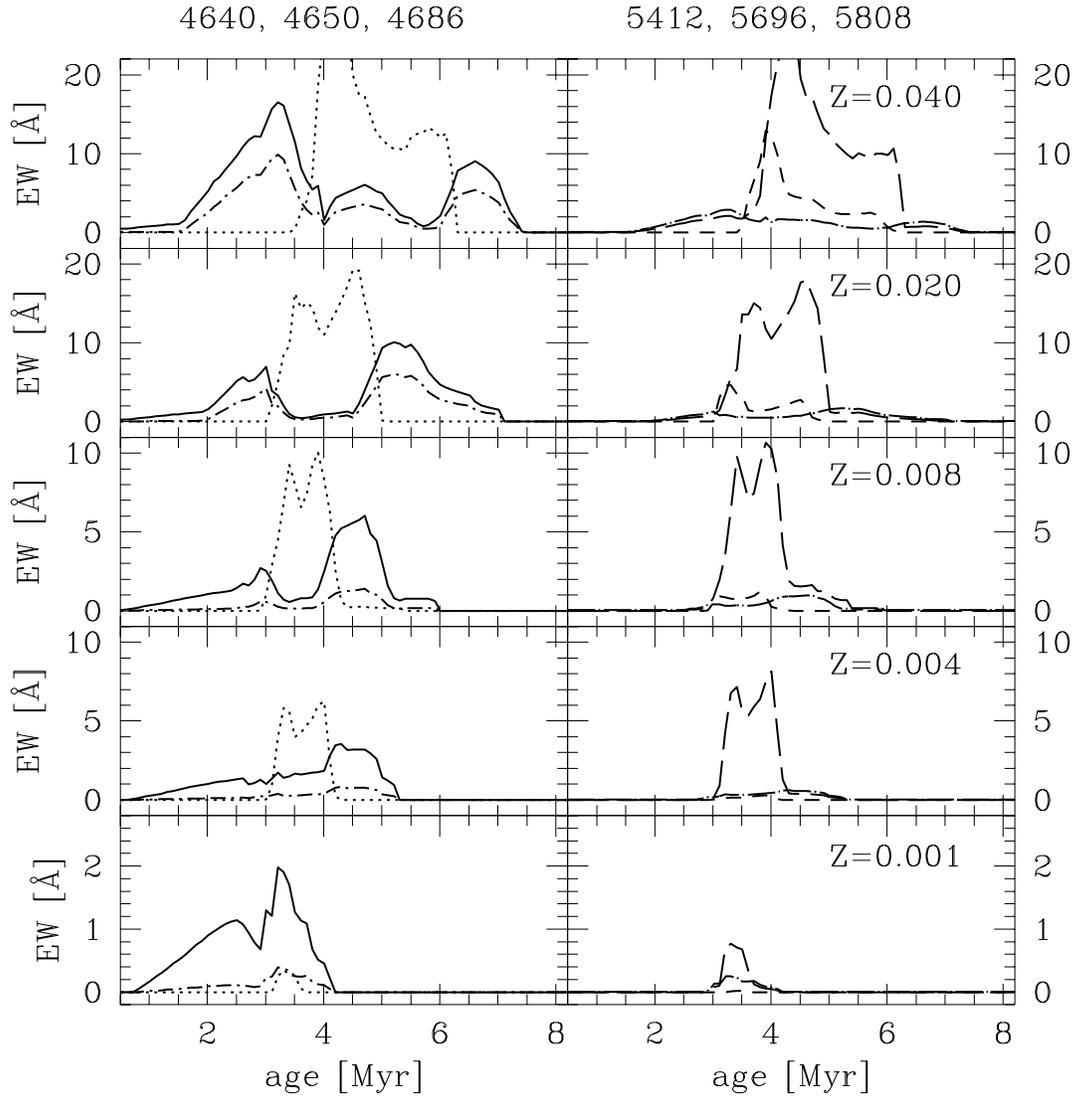,height=15cm}}
\caption{Predicted equivalent widths of optical WR lines for all 
metallicities. 
{\em Left panels:} 
\protect\heii $\lambda$4686   (solid), 
C~{\sc iii/iv} $\lambda$4650 (dotted),
N~{\sc iii/v} $\lambda$4640 (short dashed-dotted).
{\em Right panels:} 
\protect\civ\ $\lambda$5808  (long dashed),
\protect\ciii\ $\lambda$5696 (short dashed),
\protect\heii $\lambda$5412  (long dashed-dotted) 
%
\label{fig_ews_opt} }
\end{figure*}

\begin{figure}[htp]
\centerline{\psfig{figure=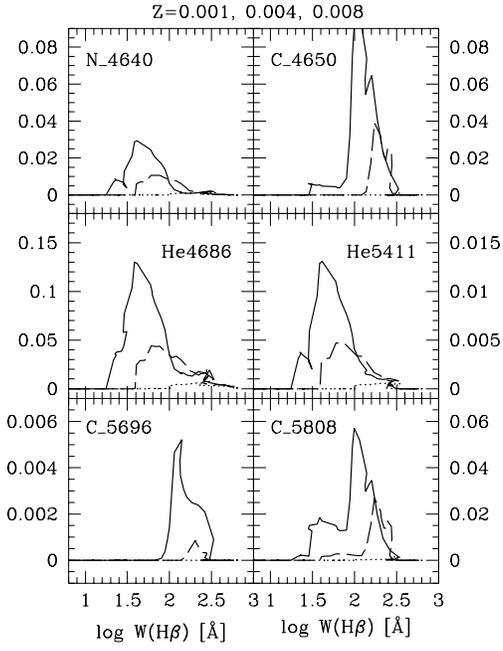,height=8.8cm}}
\caption{Predicted intensity ratio $I(\lambda)/I(\rm H_\beta)$ 
for WR lines at $Z$=0.008 (solid), 0.004 (long-dashed), and 0.001 (dotted)
\label{fig_lhb_opt} }
\end{figure}

\begin{figure}[htp]
\centerline{\psfig{figure=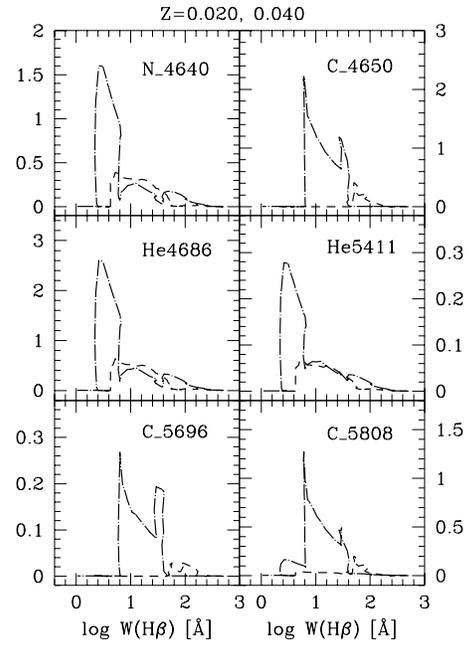,height=8.8cm}}
\caption{Same as Fig. \protect\ref{fig_lhb_opt} for $Z$=0.02 (short-dashed)
and 0.04 (dashed-dotted) 
\label{fig_lhb_opt_2} }
\end{figure}

In Figure \ref{fig_ews_opt} we plot the time evolution
of the equivalent widths of all individual WR lines (except
\halpha and \hbeta; cf.~\S\ \ref{s_hlines}).
The inclusion of 4686 \AA\ emission from Of stars, although its strength 
is considerably uncertain, leads to typical values of $W(4686) \sim$ 1-2 \AA\ 
before the appearance of the first WR stars ($t \la$ 2-3 Myr).
During the WN dominated phases the strongest lines are \heii\ $\lambda$
4686, N~{\sc iii/v} $\lambda$ 4640 and C~{\sc iv} $\lambda$ 5808 in
decreasing order.  
The weakness of N~{\sc iii/v} $\lambda$4640 at subsolar metallicity
is due to the metallicity-dependence of this line (SSM96).
For solar and higher metallicities N~{\sc iii/v} $\lambda$4640 is
expected to have a strength comparable to \heii\ $\lambda$4686
(Smith 1991, SMM96).  
In the WC-rich phase C~{\sc iii/iv} $\lambda$
4650 is expected to dominate the 4660 bump ($W(4650)/(W(4640)+W(4686))
> 1$).  Interestingly, for a given metallicity at $Z > 0.001$, the
maximum equivalent width attained for the 4650 bump, \heii\ $\lambda$
4686 and \civ\ $\lambda$5808 are all of the same order.
Since late WC stars are only expected at high metallicities 
(cf.\ \S\ \ref{s_wcwo}) the maximum equivalent width of 
\ciii\ $\lambda$ 5696 only reaches values above $\sim$ 1 \AA\ for 
$Z \ge 0.02$.
At subsolar metallicities the equivalent width of \heii\ $\lambda$5412 
is found to follow closely that of N~{\sc iii/v} $\lambda$4640.
In all cases $W(5412)$ is less than 2 \AA.

Figures \ref{fig_lhb_opt} (for $Z$=0.008, 0.004, and 0.001)
and \ref{fig_lhb_opt_2} (for $Z$=0.02 and 0.04) show the
ratios of the WR lines with respect to \hbeta\ as a function
of the \hbeta\ equivalent width.
These Figures involve purely observational quantities and can be used
for direct comparisons with integrated spectra of a young burst
population (with the aforementioned cautions in mind).
Note that for all metallicities \heii\ $\lambda$4686/\hbeta\ 
represents the maximum $WR/\hbeta$ intensity ratio obtained in a burst,
although the maximum 4650 and 5808 equivalent widths are comparable 
or even larger than $W(4686)$. This is essentially due to stronger 
\hbeta\ emission during the WC-rich phase at earlier ages, while 4686 
emission from WN stars persists until later times where \hbeta\
has already considerably decreased.

\subsubsection{Nebular and stellar 
		\protect\heii\ $\lambda$4686}
\label{s_4686}
The case of \heii\ $\lambda$4686 deserves closer examination.  As
discussed in \S\ \ref{s_nebular} the models predict a non-negligible
nebular \heii\ $\lambda$4686 emission during part of the WR-rich phase.
It is therefore expected that the 4686 \AA\ feature consists of both broad 
stellar and narrow nebular emission. Although observationally the distinction
between narrow and broad components is not necessarily straightforward
and has not generally been made, it is crucial for our understanding of
excitation conditions and WR populations especially in low metallicity
objects (cf.~Garnett \etal\ 1991, and references therein; Schaerer
1996c, Izotov \etal 1996). A detailed discussion of the origin of
nebular \heii\ and comparisons with recent observations 
will be presented elsewhere.  Here we shall present only the main
theoretical predictions.  

\begin{figure}[htp]
\centerline{\psfig{figure=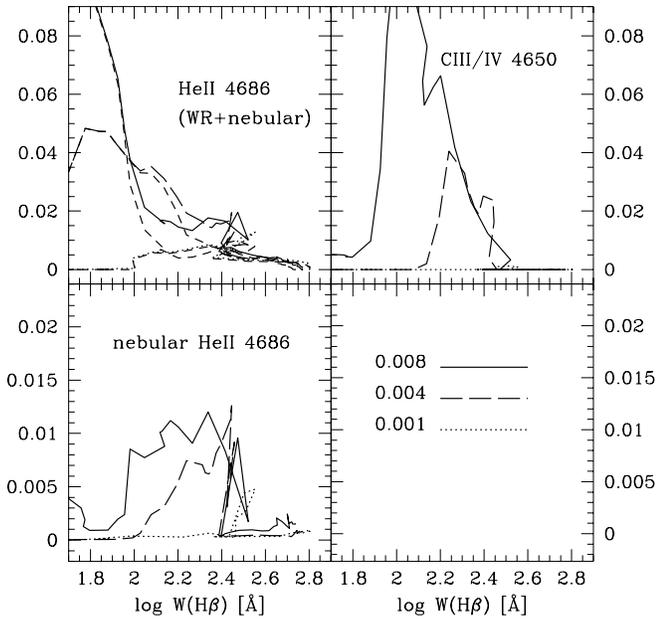,height=8.8cm}}
\caption{Predicted intensity ratios $I(\lambda)/I(\rm H_\beta)$ 
for nebular and WR lines at $\lambda \sim$ 4650-4700 \AA\ for 
Z=0.008 (solid), 0.004 (long-dashed), and 0.001 (dotted)
\label{fig_heii_lhb} }
\end{figure}

Figure \ref{fig_heii_lhb} shows the intensity ratios $\lambda$/\hbeta\ 
of pure nebular \heii\ $\lambda$4686 (lower left) and both nebular and 
total WR+nebular 
4686 emission (upper left) for metallicities $1/20 \le Z/Z_\odot \le 1/5$ 
as a function of $W(\hbeta)$. For lower \hbeta\ equivalent
widths (i.e.~larger ages) 4686 is always dominated by 
stellar emission, and is therefore not shown here. The same also holds 
for solar and higher metallicities, as discussed by Schaerer (1996c).
In the early phase of the burst nebular \heii\ 4686/\hbeta\ reaches 
values up to $\sim$ 1\%, the maximum value being nearly independent
of metallicity. 
Since nebular 4686 emission is related to WC/WO stars and their hot WN 
star precursors (cf.~\S\ \ref{s_nebular}) whose population diminishes 
with $Z$, the duration of the phase with strong nebular emission is 
also predicted to decrease.
At $Z=$0.001, the current models predict that \heii\ $\lambda$4686 
is dominated by stellar emission.

In Fig.~\ref{fig_heii_lhb} we have also plotted the predicted 
C~{\sc iii/iv} $\lambda$ 4650 emission which is due to WC/WO stars.
As mentioned ealier (\S\ \ref{s_wrinput}) this blend includes
\ciii\ $\lambda$4650, \civ\ $\lambda\lambda$4650,4658 and a small
fraction of \heii\ $\lambda$ 4686. 
Interestingly, at $Z=$ 0.008 and 0.004, the predicted 4650 emission
has a strength similar to that of the total 4686 emission.
It should be remembered, however, that at low $Z$, WC4 or WO stars 
are the dominant WC subtypes (Fig.~\ref{fig_wcsubtype}). 
In view of the strong variation of the
adopted \civ\ emission lines between WC4 and WO stars (see Table 
\ref{wc_lines}) and present uncertainties in the assignment of 
WC/WO subtypes (cf.~Smith \& Maeder 1991, Kingsburgh, Barlow \& 
Storey 1995), the predictions for 4650 remain somewhat uncertain. 

If nebular \heii\ emission is indeed related to the WC/WO phase, 
one expects to detect other signatures for these stars. 
A nearly unambigous signature from WC/WO stars is \civ\ $\lambda$5808 
whose strength is shown in Fig.~\ref{fig_lhb_opt_2}. However, the same word of 
caution regarding the strength of 4650 also applies to the strength of 5808.
The detection of \ion{O}{4} $\lambda$3400, \ion{O}{6} $\lambda$3811,34,
or \ion{O}{5} $\lambda$5590 would be a clear signature of WO stars (see
Kingsburgh, Barlow \& Storey 1995). Test calculations with the 
current models, however, predict very small equivalent widths 
for these features.

\subsubsection{Broad \protect\halpha\ and \protect\hbeta\ components}
\label{s_hlines}
The predicted emission in the \heii\ $\lambda$4861 + \hbeta\ blend
(denoted as 4861 in Tables \ref{wn_lines} and \ref{wc_lines}) and in the 
\heii\ $\lambda$6562 + \halpha\ blend (denoted as 6560) is illustrated
in Fig.\ \ref{fig_hahb} for all metallicities.
For subsolar metallicities equivalent widths of 
$\sim$ 5-25 \AA\ for 6560 and 1-3 \AA\ for 4861 are predicted. 
For solar and larger metallicities stronger emission is expected.

The line strengths predicted in this work do not include stellar
absorption features. Such predictions for young bursts have been 
given by Olofsson (1995a)
who synthesized the major H and He absorption lines for instantaneous
bursts at metallicities $Z$ $\le$ \zsun. For the ages shown in Fig.\
\ref{fig_hahb} his predicted equivalent widths (absorption) are:
2-4 \AA\ for \halpha\ and 2-5 \AA\ for \hbeta\ for the range
of IMF considered. The strength of the absorption lines is fairly 
independent of metallicity (Olofsson 1995a).
(Note, however, that for unknown reasons the EWs of the Balmer lines 
predicted by Olofsson are systematically smaller than those of 
D\'{\i}az 1988.)
The comparison with the results of Olofsson (1995a) shows that during the
WR phase the equivalent width of the \halpha\ emission from WR stars is
larger than the equivalent width of the underlying absorption. The net result 
of the superposition depends, of course, on the relative widths of the two
components.

The strength of the stellar 6560 \AA\ emission is compared to that of the 
nebular emission in Figure \ref{fig_hahb}. For low metallicities the WR emission 
may reach several percent of the nebular flux, whereas much larger contributions
are expected at solar and higher metallicity. This effect is bolstered
by both the increase of the WR population and the lower ionizing flux.
As can be seen from Table \ref{wn_lines} and \ref{wc_lines} the relative 
6560/4861 line emission in WR stars varies between 1.6 and 6.
For an integrated population a typical value of this ratio is found to be $\sim$ 4-5,
which is larger than the value of \halpha/\hbeta=2.86 for Case B 
recombination. Thus, the relative contributions of WR emission to the nebular 
\halpha\ and \hbeta\ lines may, in principle, affect reddening 
determinations derived from these lines. However, 
the uncertainties on the WR line emission in these lines are fairly
large (cf.\ Tables \ref{wn_lines}, \ref{wc_lines}).

Obviously the broad \halpha\ and \hbeta\ emission components discussed
here occur only during the WR phase, which is best detected by the presence
of WR bumps that are uncontaminated (or only weakly contaminated) by nebular emission.
In this case the flux in the broad 4861 \AA\ emission provided by WNL stars is 
typically $\sim$ 25 \% of the flux in the total 4650 bump, while it is
negligible for WC stars (cf.\ Tables \ref{wn_lines}, \ref{wc_lines}).
Additional broad emission which may be present (e.g.\ Roy et al.\ 1992,
Izotov et al.\ 1996) cannot be attributed directly to WR stars.

\section{Models including massive close binary stars}
\label{s_binary}
The main results derived from models in which binary stars are included 
according to the prescriptions in \S \ref{s_models}
are illustrated in Figures \ref{fig_binaries_008} and \ref{fig_binaries_001},
for metallicities $Z$=0.008 and 0.001. (The results for
the remaining metallicities show the same qualitative behaviour
and therefore are not be reproduced here.)
%
The top panel shows the relative fraction of WR stars of different 
subtypes. The use of WR/O number ratios (as e.g.\ shown in Fig.\
\ref{fig_wrratios}) is avoided for two reasons. First, in the phase where 
the WR stars are formed through the binary channel ($t \ga$ 5 Myr)
the number of single O stars is usually either low or possibly zero. Second,
in order to compute such ratios one also needs to account for O stars formed 
from the mass-gainer in the binary systems (whose fate is not followed
in the current calculations). In any case, as discussed earlier, 
it is advisable to directly use diagrams involving only observable 
quantities (cf.\ bottom panel).
The middle panel shows the time evolution of the equivalent widths
of the strongest WR lines (\heii\ $\lambda$4686 and \civ\ $\lambda$5808)
as a function of time. The intensities of the same lines with respect
to \hbeta\ are plotted in the bottom panel as a function of the
\hbeta\ equivalent width.

Accounting for the WR formation in massive close binary systems obviously
extends the period where WR stars are present in instantaneous bursts 
(see top panels), since their progenitors can have masses smaller than
the WR mass limit for single stars. For the binary models adopted here,
WR stars can be present up to ages of $\sim$ 11-13 Myr. 
However, this period can be even longer (ages up to $\sim$ 15-20 Myr) if the possible
evolution of the secondary to the WR phase is also included in the model (Cervi\~no, 
Mas-Hesse \& Kunth 1996, Vanbeveren, Van Bever \& Donder 1997).
In sharp contrast to burst models which include only single stars, the total 
duration of the WR phase is now essentially independent of metallicity.

The time evolution of the relative fraction of the different WR subtypes 
(top panels) is easily understood by combining the results given
in Fig.\ \ref{fig_wrratios} and Fig.\ 3 from de Loore \& Vanbeveren (1994), 
which give lifetimes in the WNL, WNE and WC phases as a function of the initial 
primary mass. The formation of 
WR stars through the binary channel extends the
final WNL phase of single stars (cf.\ \S \ref{s_pops}) to later ages,
leads to a second WC-rich period, and results in a significant
phase dominated by WNE stars after about 7-8 Myr.
However, the following caveat regarding the WR subtype distribution 
in close binary systems should be kept in mind:
As shown by Vanbeveren (1995) the predicted WC/WN number ratio for low
metallicity regions with constant star formation exceeds 
the observed values. One possible explanation is that the duration of
the WC phase predicted from binary models is too long. Indeed
the different observed WR fractions are well reproduced if RLOF 
leads to WR stars of mostly WNE type (see Maeder \& Meynet 1994).
Therefore, the relative distribution of WC and WNE subtypes in the binary phase
should be considered fairly uncertain.

The temporal evolution of the predicted equivalent widths of 4686 and 5808 
(see middle panel) shows two main characteristics:
{\em 1)} The phases where WR stars are formed predominantly through the 
single and the binary channels are well separated. This behaviour, which 
is obviously true only for short bursts, is a direct consequence
of the different progenitor masses from single and binary stars. 
{\em 2)} The expected maximum equivalent width of \heii\ $\lambda$4686 
(\civ\ $\lambda$5808)
during the binary dominated phase is $\sim$ 0.5-2 (2-4) \AA, and is fairly
independent of $Z$.
{\em 3)} At low $Z$ ($Z$ \la 0.001) the maximum equivalent widths of the
WR features during the ``binary dominated'' phase become comparable or even
stronger than those during the preceding ``single star'' phase.

The bottom panel of Figs.\ \ref{fig_binaries_008} and \ref{fig_binaries_001}
shows the predicted line ratios of 4686/\hbeta\ and 5808/\hbeta\ as a 
function of the \hbeta\ equivalent width.
As above the two phases where WR originate mostly from pure stellar wind
mass loss (``single star channel'') or from RLOF (``binary channel'')
are well separated. For $Z$ $\ge$ 0.004 ($Z$=0.001) the binary-dominated phase 
corresponds to $W(\hbeta) \la 40$~ (100) \AA.
The \hbeta\ equivalent width (or other age indicator) should thus roughly
provide a means to identify burst regions where WR formation through RLOF
may be of importance (see also Schaerer \& Vacca 1996).
Due to the strongly decreasing \hbeta\ luminosity the 
WR/\hbeta\ line ratios during the late binary phase can easily exceed
the values during the early single-star dominated phase.
The maximum line ratios of 4686/\hbeta\ and 5808/\hbeta\ obtained for 
the lowest metallicities are typically 5 to 10 \% of \hbeta\ for the
adopted binary fraction. These values might be overestimated by a factor
of $\sim$ 2 if at low $Z$ the binary frequency is actually lower than the adopted 
value (cf.\ \S \ref{s_bin_input}).

Regarding the predictions including massive close binary systems we
wish to stress again the numerous uncertainties mentioned
in \S \ref{s_bin_input}. It is certainly true that 
our current knowledge of the evolution of binary stars is less accurate
than that of single stars. In addition, in view of the simplified
treatment adopted here for binary systems we regard the detailed 
predictions as explorative and warn the reader from an overinterpretation
of this data. The general conclusions should, however, be fairly
reliable.

\section{On determining the WR and O star content in \protect\hii\ regions
and starbursts}
\label{s_wro}
The absolute number of WR and O stars, as well as the relative WR/O
star number ratios in starburst regions, have been estimated with
various techniques (e.g.~Kunth \& Sargent, 1981; Arnault, Kunth \&
Schild 1986; Leitherer, 1990; Mas-Hesse \& Kunth 1991a, VC92,
Vacca 1994; Cervi\~no \& Mas-Hesse 1994; Meynet 1995; Schaerer
1996c).  Knowledge of the absolute numbers of stars constituting a
population can be used to determine quantities such as the star
formation efficiency or rate in a region; relative WR/O star number
ratios can be used in conjunction with stellar evolution models to
constrain parameters such as the burst duration and the IMF slope, and
to understand the variation in population parameters with metallicity.

Rough estimates of the WR and O star content in a starburst region can be obtained
in a straightforward way. 
However, as demonstrated by Schaerer (1996c), more accurate techniques
are required to obtain reliable constraints on both the O star
content (cf.~Vacca 1994) and the IMF (cf.~Meynet 1995; Contini, 
Davoust \& Consid\`ere 1995).
In the following we will therefore review the various methods employed 
and discuss procedures that should yield the most accurate results.

\subsection{Determination of the number of O stars}
\label{s_no}
Usually the number of O stars present in an integrated population is
estimated from comparing {\em 1)} the total UV continuum luminosity
with the predictions of evolutionary synthesis models (e.g.~Leitherer
\& Heckman 1995), or {\em 2)} the observed strength of the H
recombination lines with predictions of the ionizing fluxes from
stellar atmosphere models (e.g.~Osterbrock 1989).  Here we shall
concentrate on the latter procedure, which has been discussed
extensively by Vacca (1994, and references therein).

\subsubsection{Using H recombination lines}
This method is based on the premise that the number of ZAMS OV stars can
be directly related to the observed \hbeta\ luminosity (cf.~Vacca
1994). The ionizing flux (in terms of the total number of H-ionizing
photons, $Q_0^{\rm Total}$) required to produce the \hbeta\ line
luminosity $L(\hbeta)$ is given by
\begin{equation}
  Q_0^{\rm Total}=
	\frac{\alpha_B({\rm H}^0) \lambda_\hbeta}
 	     {\alpha^{\rm eff}_\hbeta h c}  L(\hbeta)
\label{eq_q0_hbeta_2}
\end{equation}
(cf.~Eq.~\ref{eq_q0_hbeta_1} with $f_\gamma=1$). 
Note that Eq.~\ref{eq_q0_hbeta_2} assumes a dust-free and ionization-bounded 
\hii\ region (Case B recombination).
If the stellar population is the only source of ionizing 
photons, $Q_0^{\rm Total}$ can be expressed in terms of the number of so-called 
``equivalent'' stars of a given subtype (here we use $N^\prime_{\rm O7V}$
to denote the number of ``equivalent O7V'' stars) responsible for producing the ionizing 
luminosity,
\begin{equation}
	Q_0^{\rm Total}= N^\prime_{\rm O7V} Q_0^{O7V},
\end{equation}
where $Q_0^{O7V}$ is the Lyman continuum luminosity of an individual O7V star.
As shown by Vacca (1994) the number of ZAMS O V stars is then related to 
$N^\prime_{\rm O7V}$ by 
\begin{equation}
	\eta_0 \equiv N^\prime_{\rm O7V}/ N_{\rm OV},
\label{eq_eta0_def}
\end{equation}
where 
\begin{equation}
	\eta_0=
	  \frac{\int_{M_{\rm low}}^{M_{\rm up}} \Phi\left(M\right) 
		Q_0^{\rm ZAMS}\left(M\right) dM}
	  {Q_0^{\rm O7V} \int_{M_{\rm OV}}^{M_{\rm up}} \Phi\left(M\right) dM},
\label{eq_eta0}
\end{equation}
represents the IMF averaged ionizing Lyman continuum luminosity of a ZAMS 
population normalized to the output of one ``equivalent'' O7V star (see Vacca 1994).
In Eq.~\ref{eq_eta0}, $\Phi$ represents the IMF with the mass limits 
$M_{\rm low}$ and $M_{\rm up}$, and $Q_0^{\rm ZAMS}(M)$ is the Lyman continuum 
luminosity as a function of stellar mass for ZAMS stars. 
The number of ZAMS OV stars is thus given by
\begin{equation}
	N_{\rm OV} = \frac{Q_0^{\rm Total}}{\eta_0 \, {Q_0^{\rm O7V}}}.
\label{eq_nov}
\end{equation}
The number of ZAMS OV stars can thus be easily derived from the
observed \hbeta\ luminosity using the extensive tables of Vacca (1994)
which provide $\eta_0$ for different assumptions of the IMF slope, mass
limits and metallicity. For a Salpeter IMF and an upper mass cut-off of
120 \msun\ e.g., $\eta_0$ is found to be close to unity.

However, for comparisons with evolutionary models and other
applications it is crucial to distinguish further between ZAMS OV stars
and O stars in general.  Indeed, for instantaneous bursts the number of
ZAMS OV stars obtained from the above procedure approximates the total
number of O stars (of all luminosity classes) only for ages less than
$\sim$ 2-3 Myr, as demonstrated by Schaerer (1996c). For larger ages,
$N_{\rm OV}$ derived with the method of Vacca (1994) generally {\em
overestimates} the true number of O stars, if one does not
appropriately reduce the upper mass limit.  Nevertheless, the procedure
of Vacca (1994) can be easily generalized to account for the temporal
evolution of the total ionizing luminosity and thus to derive the total
numbers of O stars.  Accordingly, Schaerer (1996c) introduced the time
dependent value of $\eta_0$:
\begin{equation}
	\eta_0\left(t\right)=
	  \frac{\int_{M_{\rm low}}^{M_{\rm up}} \Phi\left(M\right) 
		Q_0\left(M,t\right) dM}
	  {Q_0^{\rm O7V} \int_{M_{\rm O}}^{M_{\rm up}} \Phi\left(M\right) dM},
\label{eq_eta0_t}
\end{equation}
where $Q_0\left(M,t\right)$ is the ionizing luminosity of a star of initial 
mass $M$ at time $t$. Note that for the integral in the denominator,
the identical definition for O stars is used as in the 
evolutionary synthesis models.
With Eq.~\ref{eq_eta0_t} the total number of O stars is now given by
\begin{equation}
	N_{\rm O} = \frac{Q_0^{\rm Total}}
		{\eta_0\left(t\right) \, {Q_0^{\rm O7V}}}.
\label{eq_no}
\end{equation}

The temporal evolution of $\eta_0(t)$ for our ``standard''
instantaneous burst models is shown in Fig.~\ref{fig_eta0_age} for all
metallicities considered in this work. As expected, the
ZAMS values ($\eta_0(t=0)$) agree well with the results of Vacca (1994)
using the same IMF. Fig.~\ref{fig_eta0_age} shows that $\eta_0$ remains 
fairly constant within the first
$\sim$ 2 Myr, and then decreases due to the evolution of the O star
population and the subsequent disappearance of the most massive stars
(cf.~Schaerer 1996c). For subsolar metallicities a short-lived peak of
$\eta_0$ is found at the beginning of the brief WR-rich phase ($t \sim$ 3
Myr) due to the large average Lyman continuum luminosity of WR stars at 
this time (cf.~Fig.~\ref{fig_meanq0_age}).  The second, brief
increase in $\eta_0$ seen in several of the panels occurs 
when the number of O stars (as defined by the lower limit on \teff)
drops to zero, and hence the denominator in Eq.\ \ref{eq_eta0_t}
approaches zero.  Obviously, the definition of $\eta_0$
becomes meaningless at this time; the ionizing flux of the population will be
provided by stars of B type or later and WR stars which may still be
present in some cases ($Z \ge$ 0.02).

Although more general than the formulation of Vacca (1994), the
explicit time dependence of $\eta_0(t)$ introduces two new variables
that must be determined before this quantity can be used to determine
the massive star population in a region.  In addition to the
assumptions about the IMF (slope and mainly upper mass limit) and
metallicity, the value of $\eta_0(t)$ also depends on the age of the
stellar population and the star formation history, for which we have
only considered the case of an instantaneous burst here.  In most
cases, the observed \hbeta\ equivalent width can be used to estimate
the population age (see Fig.~\ref{fig_hbeta_age}). The star formation
history is somewhat more problematic, as this is usually one of the
parameters trying to be determined.
Note that the procedures and limitations discussed here also apply 
in principle to methods which estimate the O star content from the UV continuum 
luminosity. However, these methods  tend to be less sensitive to variations
in the population parameters and avoid other potential difficulties affecting 
the nebular lines (photon leakage etc.).

\subsubsection{Using He recombination lines}
An equivalent of the above procedure can be formulated using the
\hei\ $\lambda$4471 line, which relies on $Q_1$ the total \hei\ ionizing
luminosity emitted by the stellar population(cf.~Vacca 1994). 
However, since stars with $\teff \gtrsim$ 40000 K provide enough 
photons above 24.6 eV to fully ionize helium, the ratio of the ionic 
fractions of \hei\ and \hi\ in the nebula becomes independent of \teff\
(see e.g., Doherty et al.\ 1995).
In this case \hei\ does not provide any additional information
compared to H recombination lines. 
In particular \hei\ $\lambda$4471/\hbeta\ will be  insensitive 
to variations in the IMF (both slope and upper mass limit) as long
as enough hot stars are present in the population to keep the
``equivalent'' effective temperature above $\sim$ 40000 K.

The situation is even more complicated for other \hei\ lines, e.g.\
the \hei\ $\lambda$2.06 $\mu$m line, whose predicted strength is dependent on
assumptions about the structure of the \hii\ region (cf.~Shields
1993).  Lines resulting from recombination of He$^{++}$ are rarely
detected in \hii\ regions and their origin is not fully understood
(cf.~Schaerer 1996c and references therein).  During the pure O star
phase of a burst (cf.~\S\ \ref{s_pops}) the
\heii\ $\lambda$4686/\hbeta\ is most likely too weak to be detected. In
later phases WR stars (or other poorly understood processes) dominate
the formation of nebular \heii\ lines.  See \S\ \ref{s_nebular} for
more details.

\subsection{Determination of the number of WR stars}
\label{s_nwr}
The total number of WR stars, and their relative subtype distribution, can be 
easily determined from the observed luminosity in a given broad WR 
line in the integrated spectrum of a starburst region (e.g., Osterbrock \& Cohen 1982; 
Kunth \& Sargent 1983; VC92). 
The reliability of this method obviously rests on the constancy of the line 
luminosity for all WR stars (cf.~\S\ \ref{s_wrinput}) and a correct
measurement of the desired WR component of the line, which is often a composite of
different WR lines (see \S\ \ref{s_wrinput}) or blended with nebular 
emission (e.g.~[\ion{Fe}{3}] $\lambda$4658, [\ion{Ar}{4}] 
$\lambda\lambda$4711, 4740).

\subsection{Determination of WR/O number ratios}
\label{s_nwro}
Two different methods have been developed to derive
WR/O star number ratios from optical spectra of integrated populations.
While the method of Arnault, Kunth \& Schild (1986) uses the flux in
the entire WR bump, the more recent work of VC92 
relies only on the flux in the broad \heii\ $\lambda$4686 line.
In the following we will briefly discuss and compare these two 
methods, point out their respective advantages, and present some 
improvements on these techniques.

\subsubsection{From \protect\heii\ $\lambda$4686 and \protect\hbeta}
Following VC92, we can write the ratio of the number of WR stars to the
number of reference O7V stars as
\begin{equation}
\frac{N_{\rm WR}}{N^\prime_{\rm O7V}} = 
	\left(\frac{Q_0^{\rm O7V}}{Q_0^{\rm WR}} \right)
	\left[\left(\frac{Q_0^{\rm Total}}{N_{\rm WR} Q_0^{\rm WR}}\right)
	      - 1 \right]^{-1}
\label{eq_wro7v}
\end{equation}
where $Q_0^{\rm WR}$ is the average Lyman continuum luminosity per WR star.
The total number of WR stars is given by the 
luminosity of the broad WR line,
\begin{equation}
	N_{\rm WR} = L_{\rm obs}(\lambda_{\rm WR}) /
		     L_{\rm WR}(\lambda_{\rm WR}),
\label{eq_nwr}
\end{equation}
where $\lambda_{\rm WR}$ represents one particular WR line (e.g.~\heii\
$\lambda$4686), and $L_{\rm WR}(\lambda_{\rm WR})$ is the line luminosity 
for a single WR (e.g. WN) star.
Inserting Eqs.~\ref{eq_q0_hbeta_2}, \ref{eq_eta0_t}, and \ref{eq_nwr}
in Eq.~\ref{eq_wro7v} one finds,
\begin{eqnarray}
\lefteqn{\frac{N_{\rm WR}}{N_{\rm O}} = \eta_0(t) } \nonumber \\
  	 & \left[\left(\frac{\alpha_B({\rm H}^0) \lambda_\hbeta}
 	                 {\alpha^{\rm eff}_\hbeta h c} \right)
	      \frac{L_{\rm WR}(\lambda {\rm WR})}{Q_0^{\rm O7V}}
	      \frac{L_{\rm obs}(\hbeta)}{L_{\rm obs}(\lambda_{\rm WR})}
		-  \frac{Q_0^{\rm WR}}{Q_0^{\rm O7V}} \right]^{-1}.
\label{eq_nwro}
\end{eqnarray}
For the nebular conditions assumed here (cf.\ \S\ \ref{s_synth})
$\alpha_B({\rm H}^0) \lambda_\hbeta/(\alpha^{\rm eff}_\hbeta h c) = 2.1
\times 10^{12}$ erg$^{-1}$.  In addition to the assumptions stated earlier
regarding the presence of dust and leakage of ionizing photons
(cf.\ \S\ \ref{s_no}), Equation \ref{eq_nwro} assumes that the
ionization is provided completely by the population of $N_O$ O stars
and $N_{\rm WR}$ WR stars with a Lyman continuum luminosity $Q_0^{\rm
O}$ and $Q_0^{\rm WR}$, respectively.  Furthermore it is assumed that
only stars of the generic subtype ``WR'' contribute to the WR line in
consideration.
Vacca \& Conti (1992) considered the case where only WNL stars are 
present. They adopted the values of 
$L_{\rm WNL}(\lambda4686)= 1.7 \times 10^{36}$ erg s$^{-1}$,
$Q_0^{\rm WNL}=1.70 \times 10^{49}$ s$^{-1}$, and 
$Q_0^{\rm O7V}=1.01 \times 10^{49}$ s$^{-1}$, in Eq.~15 of VC92
for $\eta_0(t)=1.$

We will now try to estimate the accuracy of the procedure of VC92, or
more generally of Eq.~\ref{eq_nwro}.  In addition to the age of the
population and the star formation history (needed for $\eta_0$;
\S\ \ref{s_no}), use of Eq.~\ref{eq_nwro} requires an estimate of
$Q_0^{\rm WR}/Q_0^{\rm O7V}$, the ratio of the average Lyman continuum
luminosity emitted per WR star to the Lyman continuum output of the
equivalent O7V star.  As shown in Fig.~\ref{fig_meanq0_age} large
variations of $Q_0^{\rm WR}$ are expected during the evolution of a
burst.  Values of $\log Q_0^{\rm WR} > 49.05$ (i.e.~$Q_0^{\rm
WR}/Q_0^{\rm O7V} > 1.68$) yield larger WR/O ratios than those derived
using Eq.~15 of VC92. However, for the typical values of
\heii\ $\lambda$4686/\hbeta $<$ 0.1 observed in starburst regions, an
average value of $Q_0^{\rm WR}$ below the value used by VC92 yields
negligible changes in $N_{\rm WR}/N_{\rm O}$.

Equation \ref{eq_nwro} accounts for the presence of one dominant WR
subtype, e.g.~WNL stars as assumed by VC92.  However, as discussed in
\S\ \ref{s_pops}, variations in the subtype distribution are expected
as a burst population ages.  Although an appropriate WR line, whose
flux is due primarily to WR stars of a certain subtype (i.e.~4686
\AA\ by WNL stars, or 5808 \AA\ by WC stars) could easily be chosen in the
case of a mixed population, for Eq.~\ref{eq_nwro} to be valid 
the representative WR subtype must also contribute the
bulk of the ionizing flux from the total WR population.

In conclusion we stress again (cf.\ Schaerer 1996c)
that variations of $\eta_0(t)$ and $Q_0^{\rm WR}$ 
complicate the derivation of accurate WR/O number ratios from 
\heii\ $\lambda$4686/\hbeta\ using the method devised by VC92.
For detailed comparisons with evolutionary models and reliable estimates of
other parameters like the IMF slope etc., it is preferable to use
models such as those developed in the present work, which allow a direct 
comparison of all relevant observational quantities 
(see Sects.~\ref{s_wrlines}) and thereby circumvent the difficulties 
discussed above.

\subsubsection{From the ``WR-bump'' and \protect\hbeta}
From the first synthesis models of WR and O star populations 
Arnault, Kunth \& Schild (1986) proposed an approximate fit
relation to derive an estimate of the total WR/O ratio (for all WR 
subtypes) from the ratio of the WR bump luminosity to \hbeta.
To the level of accuracy considered in their work the results
for a given IMF slope were found to be fairly insensitive to the 
star formation scenario and metallicity. Despite this insensitivity,
to the best of our 
knowledge the fit relation proposed by Arnault, Kunth \& Schild (1986)
has not been used in the literature to estimate WR/O ratios.
With the new models presented in this work, which rely on
more complete input physics, we have reexamined the results 
of Arnault, Kunth \& Schild (1986). 
To calculate the total emission in the WR bump, we sum the
individual contributions of N~{\sc iii/v} $\lambda$4640,
C~{\sc iii/iv} $\lambda$4650 and \heii\ $\lambda$ 4686 from
WN and WC stars.
Figure \ref{fig_arnault_wro} shows the WR/(WR+O) ratio as a function 
of the predicted WR-bump/\hbeta\ ratio for all instantaneous burst 
models (thin lines).

As anticipated from the results of Arnault, Kunth \& Schild, the data
show reasonably well-defined relation extending over at least 2 orders 
of magnitude in WR-bump/\hbeta.
Our data can be fitted by
\begin{eqnarray}
\lefteqn{\log\left[\frac{\rm WR}{({\rm WR + O})}\right] = 
	(-0.11 \pm 0.02) +} \nonumber  \\
&  (0.85 \pm 0.02) \, \log\left(\frac{\rm WRbump}{\hbeta}\right)
%
%
\label{eq_fitwro}
\end{eqnarray}
(thick solid line labeled SV) with a rms of 0.22 dex.
As discussed below (cf.~\S\ \ref{s_wrbumps}), the 
detailed predictions show notable differences if we adopt the WR bump
luminosities of Smith (1991, cf.\ Table \ref{ta_smith}). 
The average fit relation is, however, only slightly changed by
this assumption.
For comparisons the thick dotted line (labeled AKS86) shows the fit 
relation from Arnault, Kunth \& Schild (1986). Although our
detailed predictions reveal large differences with the results of
Arnault, Kunth \& Schild (1986),
the derived fit relations agree relatively well.

We have also compared the model predictions with the WR/(WR+O) ratio
derived from Eq.~\ref{eq_nwro} assuming $\eta_0(t)=1$,
$Q_0^{\rm WR}/Q_0^{\rm O7V}=1.68$ and two values of the 
WR-bump luminosity $L_{\rm WR}(\lambda {\rm WR}) = 
1.7 \times 10^{36}$ erg s$^{-1}$ (thick long-dashed line) and
$5. \times 10^{36}$ erg s$^{-1}$ (thick short-dashed line) respectively.
The long-dashed line thus corresponds to Eq.~15 of VC92, while the
short-dashed line stands for the upper limit of the WR-bump luminosity 
(WC stars from Smith 1991).
The comparison of both curves with the model predictions confirm
the finding that in general the method of VC92 {\em overestimates}
the WR/O ratio. The largest differences are obtained for large values
of WR-bump/\hbeta, which are expected at late epochs in bursts.
In this case the difference is due to the assumption of
an unevolved ZAMS population, i.e. $\eta_0=const$ (cf.~above).

We conclude that Eq.~\ref{eq_fitwro} is useful for estimates of the 
total WR/(WR+O) ratio from relatively low resolution integrated spectra of 
starburst regions.
As a word of caution it must be mentioned that such a
relation does {\it not} hold for the \heii\ $\lambda$4686/\hbeta\ 
(or even for \civ\ $\lambda$5808/\hbeta) as can be seen from
Figs.\ \ref{fig_wrratios} and \ref{fig_lhb_opt}. 
The fit relation also requires that all possible contamination
from nebular lines (cf.~\S\ \ref{s_nwr}) to the WR bump be excluded before
it can be used to determine the WR/O star number ratio. In general it
may not be possible to do this.

\subsubsection{From UV spectra}
\label{s_1640}
Since the luminosity of WN and early WC stars in the \heii\ 
$\lambda$1640 line is quite similar (see Tables \ref{wn_lines}
\ref{wc_lines}), 
and since O stars are the dominant contributors to the UV continuum, 
the equivalent width of \heii\ $\lambda$1640 is expected to be
closely related to the ratio of the total number of WR stars
to O stars.
Indeed, as Fig.~\ref{fig_new_wro} shows, our models predict 
a relatively well-defined relation between $W(1640)$ and WR/(WR+O),
which can be fitted by
\begin{eqnarray}
\lefteqn{\log\left[\frac{\rm WR}{({\rm WR + O})}\right] = 
	(-1.31 \pm 0.03) +} \nonumber  \\
&  (1.52 \pm 0.05) \, \log\left({\rm EW}(1640)\right)
\label{eq_fitwro_1640}
\end{eqnarray}
with a rms of 0.29 dex (thick solid line in Fig.~\ref{fig_new_wro}),
where $W(1640)$ denotes the equivalent width of the 
broad stellar \heii\ $\lambda$1640 emission in \AA.
Note that in this fit we have excluded the points with WR/(WR+O)=1, 
corresponding to bursts with ages $t \ga$ 3-4 Myr at $Z\ge 0.02$.

Equation \ref{eq_fitwro_1640} can be used to obtain estimates
of the total WR/O ratio from a pure equivalent width measurement
in cases where the restframe UV is the only accesible 
wavelength range (e.g.~high redshift galaxies),
or if the optical spectrum is strongly contaminated by old populations.
As long as there is no emission component from an active galactic nucleus, 
the \heii\ $\lambda$1640 line also has the advantage of being
independent of nebular lines, which may be subject to additional 
complications, as mentioned above.
For more precise determination of the WR and O star content,
comparisons with the detailed model predictions discussed in 
\S\ \ref{s_wrlines} should be preferred over the use of Eq.\ 
\ref{eq_fitwro_1640}.

\section{Summary}
\label{s_discuss}
We have constructed new evolutionary synthesis models for young
starbursts using the latest stellar evolution tracks, theoretical
spectra generated with the most complete input physics, and a
compilation of observed emission line strengths for WR stars.  To
provide predictions for various stellar and nebular emission line
strengths, our models account explicitly for the the contributions of
WR stars of different subtypes. We have also explored the consequences
of WR formation through mass transfer in close binary systems.

\subsection{Predicted observational features}
The models presented here, and their predictions for various emission
line strengths, can be used to derive 
massive star populations in starburst regions. Comparisons between 
observed and 
predicted line strengths allow the derivation of several properties 
characteristic of a starburst (e.g.\ burst age and duration, IMF, mass).
In particular, we synthesize seven different WR emission lines, including 
one UV line. We also predict the strength of the various nebular 
and stellar components contributing to the so-called WR-bump 
($\lambda \sim$ 4650 \AA).
These lines, whose emission strengths
vary considerably among the various WR subtypes, provide a means of
deriving the numbers of both WN and WC stars in a starburst region. 
The model predictions for the nebular lines provide a measure of the 
O star population. 
In addition, our models provide the first quantitative predictions
regarding a possible origin for nebular \heii\ emission which is
frequently observed in low metallicity \hii\ regions (e.g.\ Garnett et
al.\ 1991). We also estimate the contribution of WR emission to broad
components seen underlying  \halpha\ and \hbeta\ lines in the spectra
of H {\sc ii} regions (e.g.\ Roy et al.\ 1992).
Taken together these diagnostic features should
allow the most accurate determination of massive star populations in young 
bursts to date.

\subsection{Stellar evolution in different environments and tests 
of the models}
Models of massive star evolution and atmospheres have not been
extensively tested in regions outside of the solar vicinity. It is
particularly important to understand the role metallicity plays in
determining the characteristics of massive stellar atmospheres and
evolution. Our models can be used to study the evolution of massive stars in
environments which are not accessible ``locally'' (e.g.\ in very low or
very high metallicity regions), and therefore they can be used to test
the models of stellar evolution and stellar atmospheres under diverse
and extreme conditions. For example, by comparing our model predictions 
with the parameters derived
from integrated spectra of starforming clusters whose stellar content
and burst properties can be determined by other means, the effects of
metallicity and environment on massive star properties and populations
can be probed.  Although such studies have only recently begun for LMC
and SMC clusters whose content is known from stellar censuses (i.e.,
from imaging and direct counting; Vacca et al.\ 1995; Vacca \& Conti 1997), they
will serve as touchstones for evolutionary synthesis models. Similar
studies can be carried out for other star-forming regions where
individual stars can be resolved with HST. It is particularly interesting
to extend such studies to extremely metal poor galaxies (e.g., I
Zw 18, Hunter \& Thronson 1995).

There are several observational studies that can be carried out to 
test the validity of our models:
{\em 1)} Determination of the frequency of WR rich starbursts and
the frequency of WC detections among WR galaxies, as first suggested
by Meynet (1995) (see \S \ref{s_pops}).
{\em 2)} Examination of the predicted relation between nebular \heii\
$\lambda$4686 and a population of WC/WO and/or hot WN stars (cf.\
\S \ref{s_nebular}).
{\em 3)} Searches for WR signatures in old bursts, 
which may be indicative of WR stars formed through mass transfer in
close binary systems (\S \ref{s_binary}).
The first investigations along these lines have recently begun
(Schaerer 1996c, Schaerer et al.\ 1997).

Of particular interest in this regard is the analysis of the optical 
spectra of the recent samples of low metallicity extragalactic
\hii\ regions, used for the determination of the primordial He
abundance (e.g.\ Izotov, Thuan \& Lipovetsky 1997; Olive, Skillman \&
Steigman 1997). These objects are ideal for testing the evolutionary and
atmospheric models in low $Z$ environments.
This is nicely illustrated by the recent detection of WC stars 
in I Zw 18 (Izotov et al.\ 1997a, Legrand et al.\ 1997). 
Studies of stellar populations in Seyfert galaxies and AGN 
(cf.\ Heckman et al.\ 1997) may also provide new insights into 
stellar evolution in metal rich environments.

\subsection{Future applications}
With the results from such analyses of stellar populations in hand,
additional studies of the starburst phenomenon can proceed quickly. 
For example, based on the modelling of the WR and O populations in 
large numbers of starburst regions, possible variations from region 
to region in the the slope of the initial mass function for massive 
stars can be investigated (cf.\ Schaerer 1996c; Mas-Hesse \& Kunth 1997).
Also the detection-frequency of WC stars in WR galaxies could provide 
limits on the burst duration (cf.\ \S \ref{s_pops}). 
Furthermore, the ejecta from a large population of WR 
stars may have a significant impact on the chemical evolution of
the host galaxy (cf.\ Pagel et al.\ 1992; Esteban \& Peimbert 1995;
Kobulnicky et al.\ 1997). With detailed knowledge of the WR and O star
populations in starburst regions, the possible links between the massive star
content and spatial variations in metallicities can be investigated.

We envision that the present models will be used primarily to interpret the
spectra of WR galaxies. These objects provide a window on the
processes occurring during the youngest phases of a starburst
(e.g.\ VC92; Contini, Davoust \& Consid\`ere 1995; Conti 1996, 
Leitherer et al.\ 1996; Schaerer 1996c; Gonz\'alez-Delgado et al.\ 1997,
Mas-Hesse \& Kunth 1997; Schaerer et al.\ 1997).  
As such, a study of their properties has relevance to our understanding of
objects ranging from nearby H {\sc ii} regions to high redshift
galaxies.  

Future studies of starburst regions will move beyond the determination
of the stellar populations and will focus on broader aspects of the 
starburst phenomenon, such as the chronology of 
starburst events, the initial mass function and its possible dependence 
on environmental effects, mixing processes in the ISM
and the chemical evolution of young systems.
Obtaining a complete and consistent picture of a starburst region, 
incorporating both the stellar content with its 
radiative, mechanical and chemical output and the evolution of the ISM 
of the host galaxy would be a major goal of these efforts. Our models
should provide a first step toward this goal.

\acknowledgments
We thank Paul Crowther, Laurent Drissen and Alex de Koter for kindly providing 
us with measurements from their spectra. We also thank Paul for answering a
number of questions regarding WR line fluxes.
DS wishes to thank Rosa Gonz\'alez-Delgado, Yuri Izotov, 
Daniel Kunth, Enrique P\'erez, and Trinh Thuan for fruitful discussions. 
Thierry Contini, Claus Leitherer and Georges Meynet also provided comments
on a earlier version of the ma\-nu\-script.
DS is supported by the Swiss National Foundation of Scientific Research
and acknowlegdes partial support from the Directors Discretionary 
Research Fund of the STScI.
WDV acknowledges support in the form of a fellowship from the 
Beatrice Watson Parrent Foundation.

\appendix
\section{Appendix}
The results from the models presented in this work are available
in electronic format. Additional data such as spectral energy 
distributions and integrated colors are also provided.
Further variations of the major input parameters (IMF slope etc.)
are also considered. The models are found on the Web at \\
{\tt http://www.stsci.edu/ftp/science/starburst }.


\clearpage


\clearpage
\clearpage

\clearpage
\clearpage

\begin{figure}[htp]
\centerline{\psfig{figure=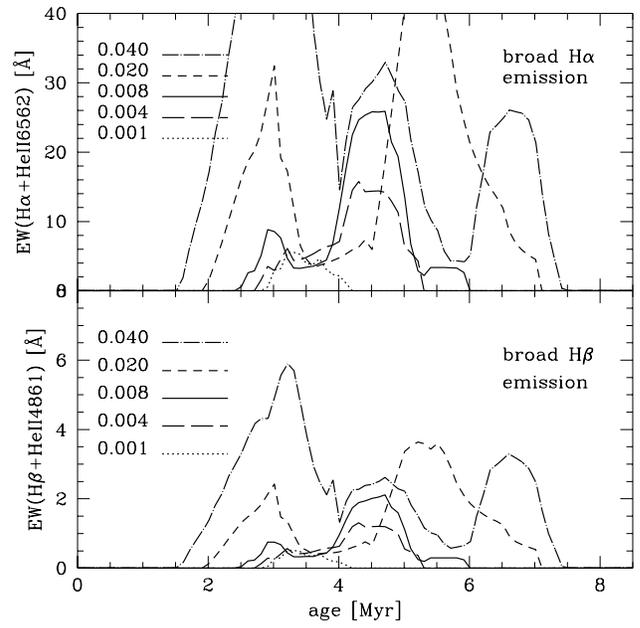,height=8.8cm}}
\caption{Predicted equivalent widths of \protect\halpha\ (upper panel) 
and \protect\hbeta\ (lower panel) blends for all metallicities. 
Symbols as in Fig.~\protect\ref{fig_wcsubtype}
\label{fig_ews_hahb} }
\end{figure}

\begin{figure}[htp]
\centerline{\psfig{figure=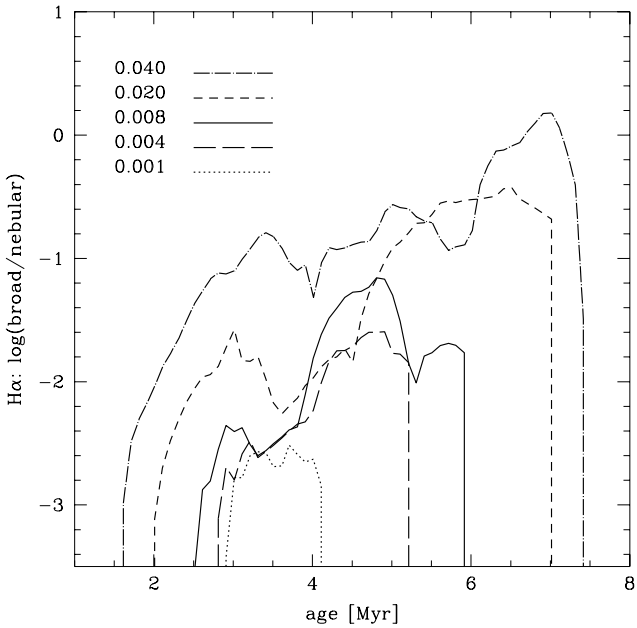,height=8.8cm}}
\caption{Ratio of broad WR emission to nebular 
\protect\halpha\ line for all metallicities.
Symbols as in Fig.~\protect\ref{fig_wcsubtype}. The relative 
contribution to \protect\hbeta\ is nearly identical to the \protect\halpha\
contribution (see \S\ \protect\ref{s_hlines})
\label{fig_hahb} }
\end{figure}

\begin{figure}[htp]
\centerline{\psfig{figure=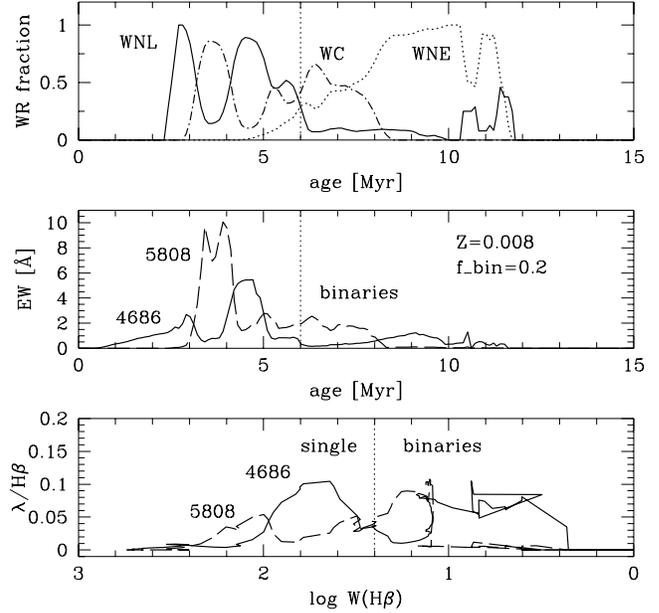,height=8.8cm}}
\caption{Model predictions including massive close binary stars with 
a binary fraction $f=0.2$ for $Z$=0.008 (instantaneous burst).
{\em Top panel:} Relative fraction of WR stars of different subtypes.
{\em Middle:} Temporal evolution of predicted equivalent widths for
\protect\heii\ $\lambda$4686 (solid) and \protect\civ\ $\lambda$5808 
(long-dashed). The arrow indicates the beginning of the phase where WR 
stars are predominantly formed through the binary channel.
{\em Lower panel:} Line ratios $\lambda$/\protect\hbeta\ as 
a function $W(\hbeta)$. Same symbols as middle panel.
The vertical dotted lines illustrates the clear separation of the 
``single star'' and ``binary'' phases in this diagram
\label{fig_binaries_008} }
\end{figure}

\begin{figure}[htp]
\centerline{\psfig{figure=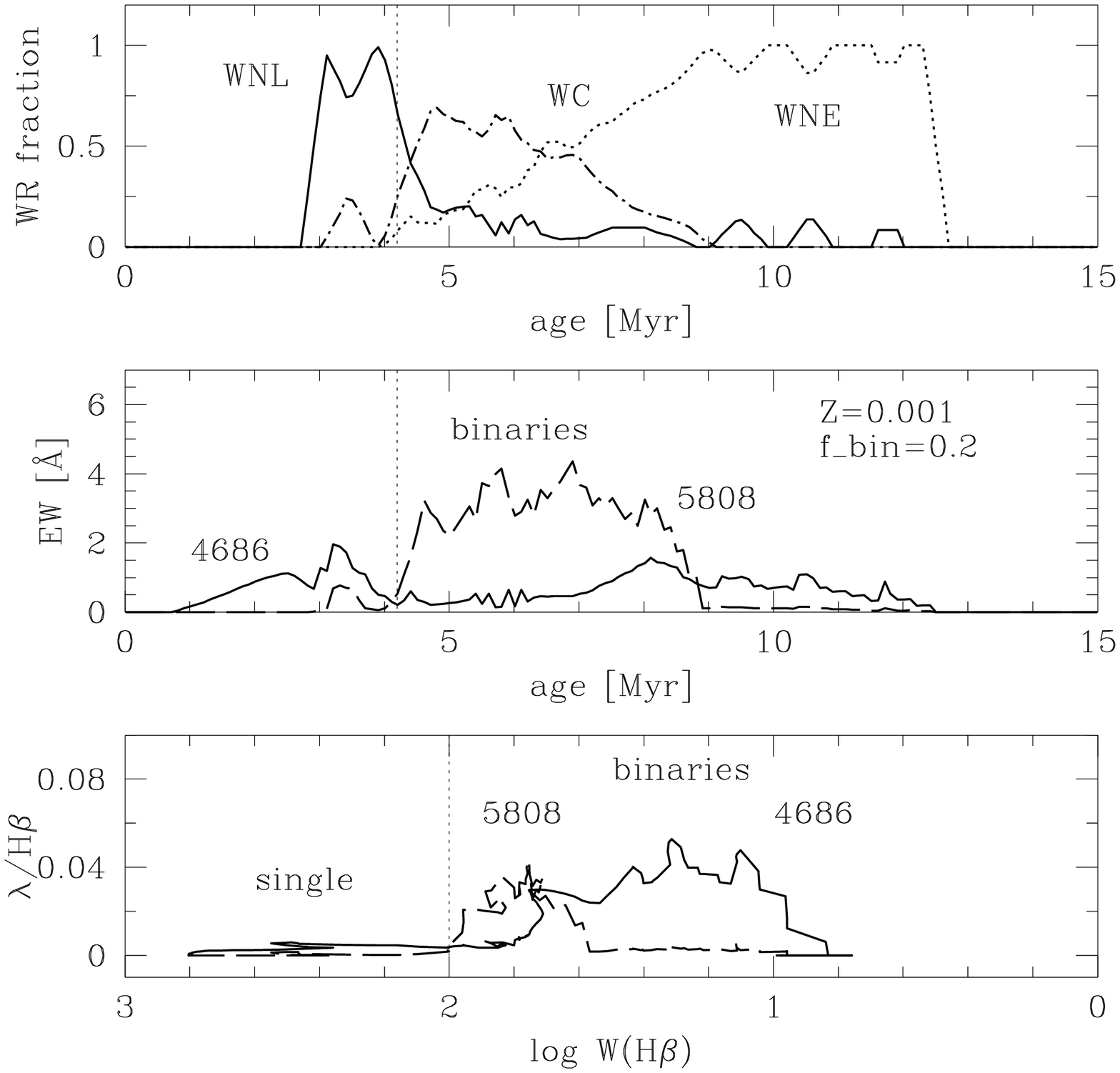,height=8.8cm}}
\caption{Same as \protect\ref{fig_binaries_008} for $Z$=0.001
\label{fig_binaries_001} }
\end{figure}

\begin{figure}[htp]
\centerline{\psfig{figure=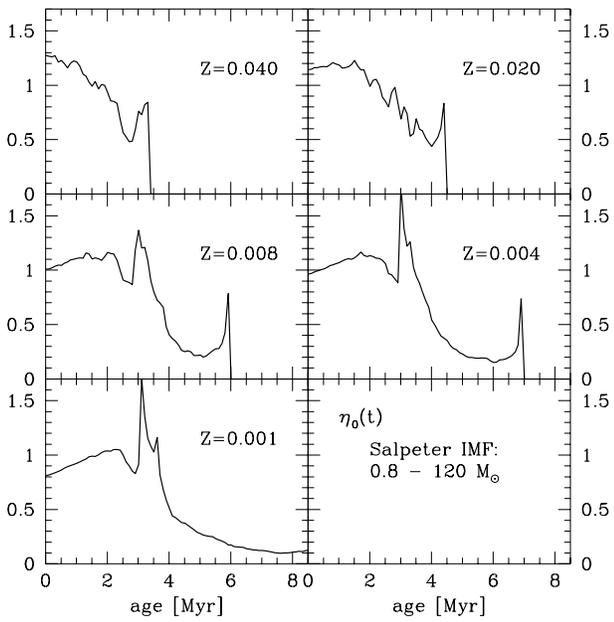,height=8.8cm}}
\caption{Evolution of $\eta_0(t)$ (defined by Eq.\ \protect\ref{eq_eta0_t})
in standard burst models for metallicities
from $Z$=0.040 to 0.001
\label{fig_eta0_age} }
\end{figure}

\clearpage
\begin{figure}[htp]
\centerline{\psfig{figure=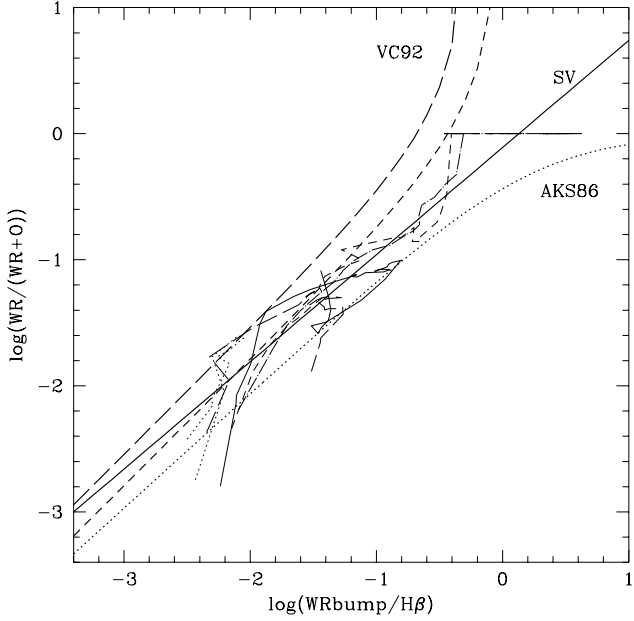,height=8.8cm}}
\caption{Number ratio WR/(WR+O) as a function of the intensity ratio
of the WR-bump over \protect\hbeta\ obtained from instantaneous burst models 
at all metallicities (thin lines using same symbols as in 
Fig.~\protect\ref{fig_wcsubtype}).
The thick solid line (SV) shows the fit to our models 
(cf.~Eq.~\protect\ref{eq_fitwro}). 
Thick dashed: relation from Arnault, Kunth \& Schild.
Thick long-dashed: relation from VC92 (their Eq.~15 assuming $\eta=1.$).
Thick short-dashed: same as long-dashed but assuming a larger average line 
luminosity of $\log$(4650-4686)=36.7
\label{fig_arnault_wro} }
\end{figure}

\begin{figure}[htp]
\centerline{\psfig{figure=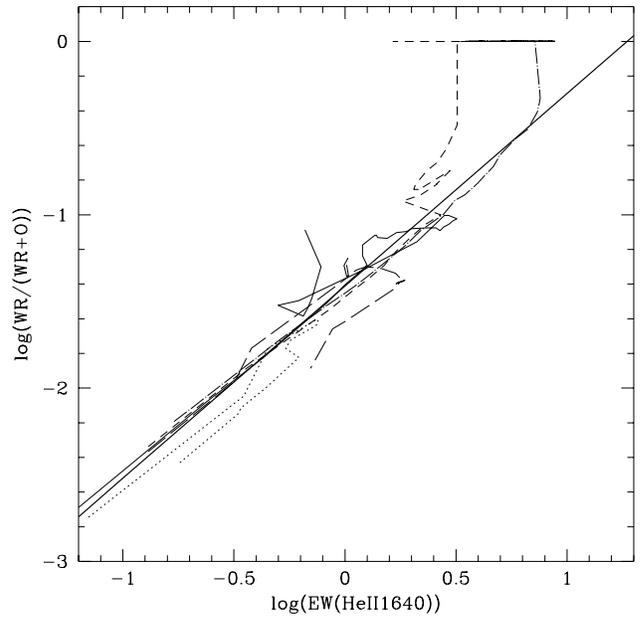,height=8.8cm}}
\caption{Number ratio WR/(WR+O) as a function of the \protect\heii\
$\lambda$1640 equivalent width (emission, in \AA)
obtained from instantaneous burst models at all metallicities 
(thin lines using same symbols as in Fig.~\protect\ref{fig_wcsubtype}).
The thick solid line shows the fit to our models 
(cf.~Eq.~\protect\ref{eq_fitwro_1640})
\label{fig_new_wro} }
\end{figure}

\clearpage
\end{document}